\documentclass[a4paper,12pt,twoside]{report}

\usepackage{graphicx}
\usepackage{verbatim}
\usepackage{latexsym}
\usepackage{mathchars}
\usepackage{setspace}
\usepackage{amsfonts}
\UseRawInputEncoding
\usepackage[left=2cm,right=2cm,top=2cm,bottom=3cm]{geometry}

\input{blocked.sty}
\input{uhead.sty}
\input{boxit.sty}
\input{icthesis.sty}

\newcommand{\NN}{{\sf I\kern-0.14emN}}   % Natural numbers
\newcommand{\ZZ}{{\sf Z\kern-0.45emZ}}   % Integers
\newcommand{\QQQ}{{\sf C\kern-0.48emQ}}   % Rational numbers
\newcommand{\RR}{{\sf I\kern-0.14emR}}   % Real numbers

\newcommand{\normallinespacing}{\renewcommand{\baselinestretch}{1.5} \normalsize}

\newcommand{\syncc}{~\stackrel{\textstyle \rhd\kern-0.57em\lhd}{\scriptstyle L}~}

\begin{document}

\title{\LARGE {\bf Scoliosis Detection using Deep Neural Network}}

\author{Nguyen Hoang Yen}
\submitdate{September 2022}
\pagebreak

\normallinespacing
\maketitle

\preface
\cleardoublepage
\addcontentsline{toc}{chapter}{Abstract}

\begin{abstract}

 %It dramatically affects the quality of life, which can cause complications from heart and lung injuries in severe cases. In current clinical settings, the severity of scoliosis is evaluated by assessing the contralateral blending angle of the spinal cord. Cobb angle is widely used in the diagnosis and treatment of scoliosis. Nevertheless, the manual measurement of Cobb angle in clinical practice is time consuming and unreliable. 

Scoliosis is a sideways curvature of the spine that most often is diagnosed among young teenagers. It dramatically affects the quality of life, which can cause complications from heart and lung injuries in severe cases. The current gold standard to detect and estimate scoliosis is to manually examine the spinal anterior-posterior X-ray images. This process is time-consuming, observer-dependent, and has high inter-rater variability. Consequently, there has been increasing interest in automatic scoliosis estimation from spinal X-ray images, and the development of deep learning has shown amazing achievements in automatic spinal curvature estimation. The main target of this thesis is to review the fundamental concepts of deep learning, analyze how deep learning is applied to detect spinal curvature, explore the practical deep learning-based models that have been employed. It aims to improve the accuracy of scoliosis detection and implement the most successful one for automated Cobb angle prediction.

%enhance the performance of automated Cobb angle prediction as well as implement Seg4Reg Network - the state-of-the-art neural network that have achieved the $1^{st}$ place in the accurate automated spinal curvature estimation (AASCE) challenge 2019.


%The main goal of this work is to study how Deep Learning is applied to the evaluation of spinal curvature as well as investigation state-of-the-art automated spinal curvature estimation algorithms.

%which sparked interest in development of accurate automated spinal curvature estimation in . paper1 Therefore, accurate computer-assisted spinal curvature measurement is necessary.
%Recent advances in deep learning show the advantages of automatic spinal curvature
%assessment. (Automated Spinal Curvature Assessment from X-Ray Images Using Landmarks Estimation Network via Rotation Proposals)
%Therefore, it’s likely essential to have a reliable estimation of Cobb angles
%The task of this challenge is to automatically obtain 3 Cobb angles from each spinal anterior-posterior X-Ray image. 
% Recently, deep neural networks have got amazing achievements in various image classification tasks. How to apply these deep models to the problem of spinal curvature estimation becomes a hot issue in automated AIS assessment.
\end{abstract}
\cleardoublepage

\addcontentsline{toc}{chapter}{Acknowledgements}

\begin{acknowledgements}

\begin{itemize}
 \item First and foremost, I would like to acknowledge and give my warmest thanks to my supervisor Dr Anh Nguyen for the continuous support of my internship at the Computer Science department of University of Liverpool, and for his patience, motivation, enthusiasm, and knowledge. His guidance and advice carried me through all the stages of writing my thesis. Dr Anh Nguyen gave me true meaning of Mathematics and Deep Learning.
 \vspace*{3mm}
 \item I would also like to thank my second advisor, Dr Khoi Le. It is indeed my privilege to work with him, who is always supportive, providing me with guidance and counsel whenever I need. His dynamism, vision, sincerity and motivation have deeply inspired me. I am grateful for what he has offered me during research work and thesis preparation.
 \vspace*{3mm}
 \item Besides my supervisors, I would like to thank the rest of my thesis committee: Dr. Aggelos K.Katsaggelos, Dr. Thrasyvoulos N.Pappas, and Dr. Son Doan, for their encouragement, insightful comments, and hard questions. Their kind help and support have made my study and life in Liverpool a wonderful time.
 \vspace*{3mm}
 \item Last but not least, I am truely grateful to my parents for their love, prayers, caring and sacrifices for educating and preparing me for my future. Also I express my thanks to my brother, sister in law and my friends for their support and valuable prayers.
\end{itemize}

\end{acknowledgements}
\cleardoublepage
\addcontentsline{toc}{chapter}{List Of Abbreviations}

\begin{ListOfAbbreviations}

\begin{table}[!htp]
\centering
\begin{tabular}{{ll}}
AASCE &	\hspace{15mm} Accurate Automated Spinal Curvature Estimation\\ 
Adam & \hspace{15mm} Adaptive Moment Estimation\\
AE & \hspace{15mm}	Autoencoders\\
AIS & \hspace{15mm}	Adolescent Idiopathic Scoliosis \\
ANN   &	\hspace{15mm} Artificial Neural Network\\
ASD	& \hspace{15mm} Average Symmetric surface Distance\\
AUC	& \hspace{15mm} Area Under the Curve\\
BG & \hspace{15mm}	Background \\
BP & \hspace{15mm}	Backpropagation\\
CNN & \hspace{15mm}	Convolutional Neural Network\\
CONV &  \hspace{15mm} layer Convolutional Layer \\
CPN	& \hspace{15mm} Cascade Pyramid Network\\
CT	& \hspace{15mm} Computed Tomography\\
DenseNet & \hspace{15mm}	Densely Connected Convolutional Neural Network\\ 
DL & \hspace{15mm}	Deep Leaning\\
ERM	& \hspace{15mm} Empirical Risk Minimization\\
FC layer  &	 \hspace{15mm} Fully-Connected Layer\\
FCN  &	\hspace{15mm} Fully Convolutional Neural Network\\
FPN	& \hspace{15mm} Feature Pyramid Network\\
GAN &	\hspace{15mm} Generative Adversarial Network\\
GPUs  &	\hspace{15mm} Graphical Processing Units\\
GRU	& \hspace{15mm} Gated Recurrent Uni\\
LSTM &	\hspace{15mm} Long-short Term Memory\\
LV &	\hspace{15mm} Left Ventricle Cavity\\
MAE & \hspace{15mm}	Mean Absolute Error\\ 
MR	& \hspace{15mm} Magnetic Resonance\\
MRI	& \hspace{15mm} Magnetic Resonance Imaging\\
MSE	& \hspace{15mm} Mean Squared Error\\
MVC-Net &	\hspace{15mm} Multi-View Correlation Network\\
MVE-Net & \hspace{15mm}	Multi-View Extrapolation Net\\
MYO  &	\hspace{15mm} Left Ventricular Myocardium\\
PSPNet & \hspace{15mm}	Pyramid Scene Parsing Network\\
RC & \hspace{15mm}	Residual Corrector\\
ReLU  & \hspace{15mm}	Rectified Linear Unit \\
ResNet & \hspace{15mm}	Residual Neural Network\\
RNN & \hspace{15mm}	Recurrent Neural Network\\
ROI	& \hspace{15mm}  Rotational region Of Interest \\
RV 	& \hspace{15mm} Right Ventricle Cavity\\
SGD	& \hspace{15mm} Stochastic Gradient Descent\\
SMAPE	& \hspace{15mm} Symmetric Mean Absolute Percentage\\
TPUs &	\hspace{15mm} Tensor Processing Units\\

 \end{tabular}

 \label{table:abbreviations}
\end{table}

\end{ListOfAbbreviations}

\body
\chapter{Introduction}

\section{Motivation and Objectives}

\subsection{Motivation}

Mathematics has been commonly applied to various fields such as physics, engineering, industry, medicine, biology, finance, business, and computer science. In the past, practical applications is closely linked with research in pure mathematics such as differential equations, approximation theory and applied probability. Today, the term ''applied mathematics'' is employed in a broader sense. It not only consists of the classical areas mentioned above but also other areas that have become gradually important in applications. For example, the four critical concepts of Mathematics including Statistics, Linear Algebra, Probability, and Calculus have contributed strongly to the generation of algorithms that can learn from data to make accurate predictions. As a result, applied Mathematics has motivated the robust development of Machine learning and especially deep learning.

% have encouraged the development of mathematical theories, which then became the subject of study in pure mathematics where abstract concepts are studied for their own sake. As a result, the activity of applied mathematics

%Applied mathematics is thus defined as a combination of mathematical science and specialized knowledge. 

Since the early 2000s, deep learning (DL) has achieved significant popularity both in the research and industry community and has been developed as a increasingly state-of-the-art technique in different areas, consisting computer vision, natural language processing, and healthcare. Generally, DL algorithms use neural networks to automatically extract a set of intricate hierarchical features from data. These features are useful for pattern recognition, decision-making, and inference as they discover the intricate structure in large raw data. This ability is highly desirable in the field of medical data analysis, as it enables one to automatically extract, analyze, and describe information from medical imaging data. Especially, neural networks can be employed to execute monotonous tasks like detection, segmentation and performing regression estimation from medical images (e.g., vertebrae detection, landmark detection, spinal curvature estimation)~\cite{10.1007/978-3-030-39752-4_9}. 

 \begin{center}
    \begin{figure}[htp]
    \begin{center}
     \includegraphics[scale=1.4]{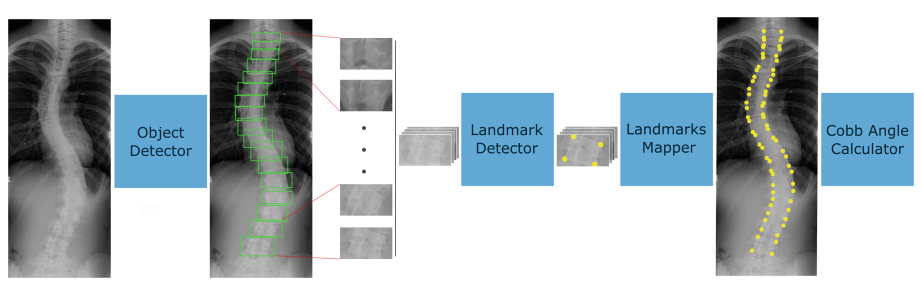}
    \end{center}
    \caption{ An overview of the pipeline}
    \label{fig1}
    \end{figure}
\end{center}

Fig.\ref{fig1} illustrates an overview of a typical spinal scoliosis measurement task in one of the most widely employed modalities where deep learning methods have been applied to. This method has a creative structure that first detects vertebrae as objects followed by a landmark detector that estimates the 4 landmark corners of each vertebra separately. The spinal curvature is then computed adopting the slope of each vertebra acquired from the predicted landmarks. This shows the common applicability of DL to a typical image task. 

Beside, with the support of advanced hardware such as graphical processing units (GPUs) and tensor processing units (TPUs), neural networks can execute prediction very fast (e.g., less than a second). Subsequently, they can be ideally utilized to decrease physicians, clinicians, and radiologists’ workload as well as improve the quality of healthcare better. 

%Since 2015, neural networks has gained significant precision and speed in many vision tasks and thus, have become the leading technique for automated medical image analysis such as anatomical structure segmentation, landmark detection, lesion detection, and segmentation, as well as image registration, image reconstruction, and computer-aided diagnosis/prognosis [2, 3]

 %\begin{center}
   % \begin{figure}[htp]
    %\begin{center}
    % \includegraphics[scale=1.4]{introduction/fig2.png}
    %\end{center}
    %\caption{ Illustrative diagram of the distributional shift between the training and testing data in real-world applications.}
    %\label{fig2}
    %\end{figure}
%\end{center}

\subsection{Objectives}

Adolescent idiopathic scoliosis (AIS) refers to the abnormal spinal curvature, which begins from later childhood or adolescence and continues to adulthood. AIS is one of the most popular type of scoliosis, with about $4\%$ apperance in adolescents~\cite{ijerph18158152}. It causes skeletal muscle dysfunction which lead to lower back pain, ventilatory restrictions, or even pulmonary cardiac failure~\cite{2}. Fortunately, it is preventable, and early diagnosis and proper intervention in adolescents are vital for controlling abnormal spinal curve progression.

Since patients with AIS may feel mild or no pain in the early stage, the diagnosis of AIS are depending on curvature assessment on lateral spinal X-ray images, and Cobb angle is one of the commonly applied standards in clinical practice. However, the manual measurement of Cobb angles is time consuming, observer-dependent and labor intensive. In particular, the anatomical variations and low contrast of X-ray images can be challenge for experienced clinicians, leading to high inter-observer variability that could negatively impact assessing prognosis and treatment decisions. Consequently, there has been increasing interest in automatic estimation of Cobb angles directly from the X-ray images. 

There have been a lot of proposed techniques for evaluation of spinal curvature recently. Anitha et al.~\cite{22,3} employed active contouring and filtering to locate the desired vertebrae and segment them, then calculated the Cobb angle based on the segmentation results. Li et al.~\cite{10.1007/978-3-319-66182-7_15} estimated vertebral
coordinates using BoostNet based on deep learning. Wu et al.~\cite{WuH} associated the multi-view X-ray features with a multi-view convolutional layer to form a Multi-View Correlation Network (MVC-Net) to compute the Cobb angle. Wang et al.~\cite{Wang2019AccurateAC} directly predicted the Cobb angle using a Multi-View Extrapolation Net (MVE-Net). Chen et al.~\cite{Chen2019AnAA} used an Adaptive Error Correction Net (AEC-Net) to convert the estimation of spinal Cobb into a high-precision regression task. Presently, the accurate automated spinal curvature estimation (AASCE) challenge 2019 organizes a competition to motivate the development of more accuracy spinal curvature estimation algorithms which automatically obtain 3 Cobb angles from each spinal anterior-posterior X-Ray image. 

The main goal of this thesis is to study how Deep Learning is applied to the evaluation of spinal curvature and investigate the state-of-the-art automated spinal curvature estimation models that have proposed in the AASCE challenge 2019 as well as implement the challenge's most successful technique in order to see the application of Mathematics in DL and gain useful experience for my future study.

\section{Contributions}
In this thesis, I concentrate on analyzing recent successful techniques which have been used to improve incredibly the generalization and robustness of neural networks. Especially, I will study the state-of-the-art models that are applied to spinal anterior-posterior X-Ray images available at SpineWeb in the Accurate Automated Spinal Curvature Estimation (AASCE) challenge 2019. Particularly, I focus on:

\begin{itemize}
    \item Review key concepts of DL, which includes studying how deep neural networks are built in details, how they are trained and what type of deep neural networks have been commonly used to accomplish the best accuracy in prediction tasks of medical images.
    \item Investigate the applications of DL in spinal imaging and especially in how to automatically predict cobb Angle directly from the X-ray scan with provided landmarks. The main concept of this part is to analyze the contribution of the 5 top-ranked teams in Spinal Curvature Estimation Challenge 2019 in order to see how their advanced models outperform other competitors.
    \item Have a deep look at the performance of the best team (Team X) in Spinal Curvature Estimation Challenge 2019~\cite{10.1007/978-3-030-39752-4_7}, which includes exploring the novel and state-of-the-art pipeline provided and how to train and test its performance on test sets as well as studying what advantages there are in this method. 
    \item Implement their submitted source code available on github~\cite{lin2019seg4reg}.
\end{itemize}

In Chapter 2, I will introduce DL basics, common existing Deep neural networks which have helped to gain the prediction accuracy as well as the training of those Deep neural networks. Next, I will review related work consisting of DL for spinal imaging and Spinal Curvature Estimation Challenge 2019 in Chapter 3. All these form the basis of the work presented in Chapters 4, which analyses the successful application of Seg4Reg network for the automated estimation of spinal curvature and implement this method. In Chapter 5, I will finally summarize the thesis's achievements, applications and future works.

% \section{Statement of Originality}

% Statement here.

% \section{Publications}

% Publications here.

\chapter{Background Theory}

A deep neural network can be defined as an artificial neural network (ANN) with an increased number of hidden layers between the input and the output layers ($n>2$), which enables itself to model complex non-linear relationships in data. Deep neural networks have been shown to be successful in several vision tasks, such as image classification~\cite{NIPS2012_c399862d,zhu2019multi}, object detection~\cite{rao2017mobile,shafiee2017fast}, semantic segmentation~\cite{nguyen2020end,bertasius2017convolutional}, optical flow estimation~\cite{revaud2015epicflow,ranjan2017optical}, medical visual question and answering~\cite{do2021multiple}, and deformation registration~\cite{tran2022light}. From the simple deep neural networks, researchers have been developing more robust and effective ones to achieve higher accuracy in diverse image tasks.

 In this chapter, we will first review the main building blocks of deep neural networks. We will then study several popular deep learning networks that have been widely utilised in the latest DL-based medical imaging applications. We finally analyse how neural networks will be trained with successful key techniques to learn proper values of the network parameters that assists in the purposes of prediction.

\section{ Core building blocks in neural networks }

In this part, we will present the main individual layers/core building blocks in neural networks. 

\subsection{Convolution Layers} 
    
   A convolutional layer (CONV layer) is the core building block of a Convolutional Network that does most of the computational heavy lifting. It contains a set of filters/kernels whose weights and biases are learnable. Each filter in a convolutional layer is small spatially (along width and height), and it is only connected to a local region of the input volume each time. A convolution layer extracts a set of feature maps (activation maps) by letting each of its filters slides across the whole input volume along the width and height and computing the dot product between each filter's weights and the input volume plus bias offsets. These activation maps correspond to the response of the convolutional filters at each spatial position of the input. For instance, given a convolutional layer with $k_{out}$ 2D $n\times n$ convolution kernels and an input image $x_{in} \in  \mathbb{R}^{H\times W\times k_{in}}$ we can formulate the computation as:
    \begin{equation}
        \forall i \in (1, k_{out}), \textbf{x}_{out}^{(i)}=\textbf{w}^{(i)} \circ \textbf{x}_{in}+b^{(i)},
    \end{equation}
    where $\textbf{w}^{(i)} \in \mathbb{R}^{n\times n\times k_{in}}$, $b^{(i)}\in \mathbb{R}$ represent the weights and bias parameters in the $i^{th}$ convolution kernel respectively, $\circ$ represents the convolution operation (i.e. dot products between the filters and local regions of the input), $x_{out} \in \mathbb{R}^{H^{\prime}\times W^{\prime}\times k_{out}}$ represents the output feature maps. $H^{\prime}$, $W^{\prime}$ are determined by the size of the kernel $n$, the stride $s$, the amount of zero padding $p$ and the input height $H$ and width $W$ respectively: 
    \begin{equation}
   H^{\prime} = (H - n + 2p)/s + 1, \hspace{1.5cm} W^{\prime} = (W - n + 2p)/s + 1
     \end{equation}
    
     Intuitively, the network will learn filters that activate when they see some kind of visual feature such as an edge of some orientation or a blotch of some color on the first layer, or finally entire honeycomb or wheel-like patterns on higher layers of the network. Additionally, if an entire set of filters in each CONV layer is 12 filters for example, each of them will produce a separate 2-dimensional activation map. These 12 activation maps will then be stacked along the depth dimension to produce the output volume which will be passed to the next layer.
   
    \begin{center}
    \begin{figure}[htp]
    \begin{center}
     \includegraphics[scale=1.3]{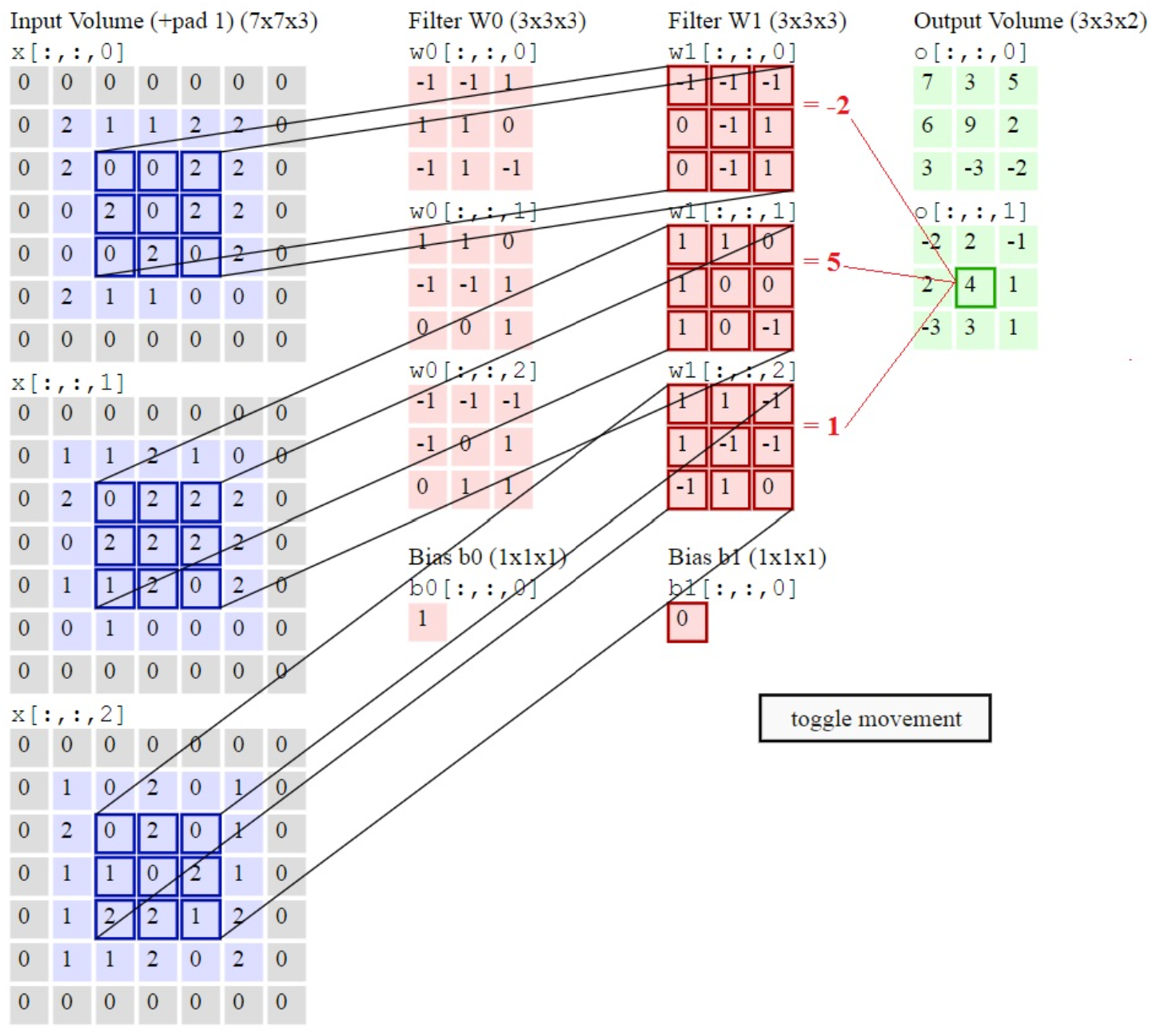}
    \end{center}
    \caption{ A demo of a CONV layer} 
    \label{fig2.13}
    \end{figure}
    \end{center}
     
Fig.\ref{fig2.13} illustrates a demo of a CONV layer. As 3D volumes are difficult to visualize, all the volumes (the input volume (in blue), the weight volumes (in red), the output volume (in green)) are visualized with each depth slice stacked in rows. The input volume is of size $(W_1 = 5, H_1 = 5, D_1 = 3)$, and the CONV layer parameters are $(K = 2, F = 3, S = 2, P = 1)$. That is, we employ two filters of size $(3 \times 3)$, and they are applied with a stride of 2. Consequently, the output volume size has spatial size $(5 - 3 + 2)/2 + 1 = 3$. Additionally, a padding of $(P = 1)$ is applied to the input volume, making the outer border of the input volume zero. The visualization iterates over the output activations (green), and illustrates that each element is calculated by elementwise multiplying the highlighted input (blue) with the filter (red), adding it up, and then offsetting the outcome by the bias.

\subsection{Activation Layers}

    \begin{center}
    \begin{figure}[htp]
    \begin{center}
     \includegraphics[scale=0.6]{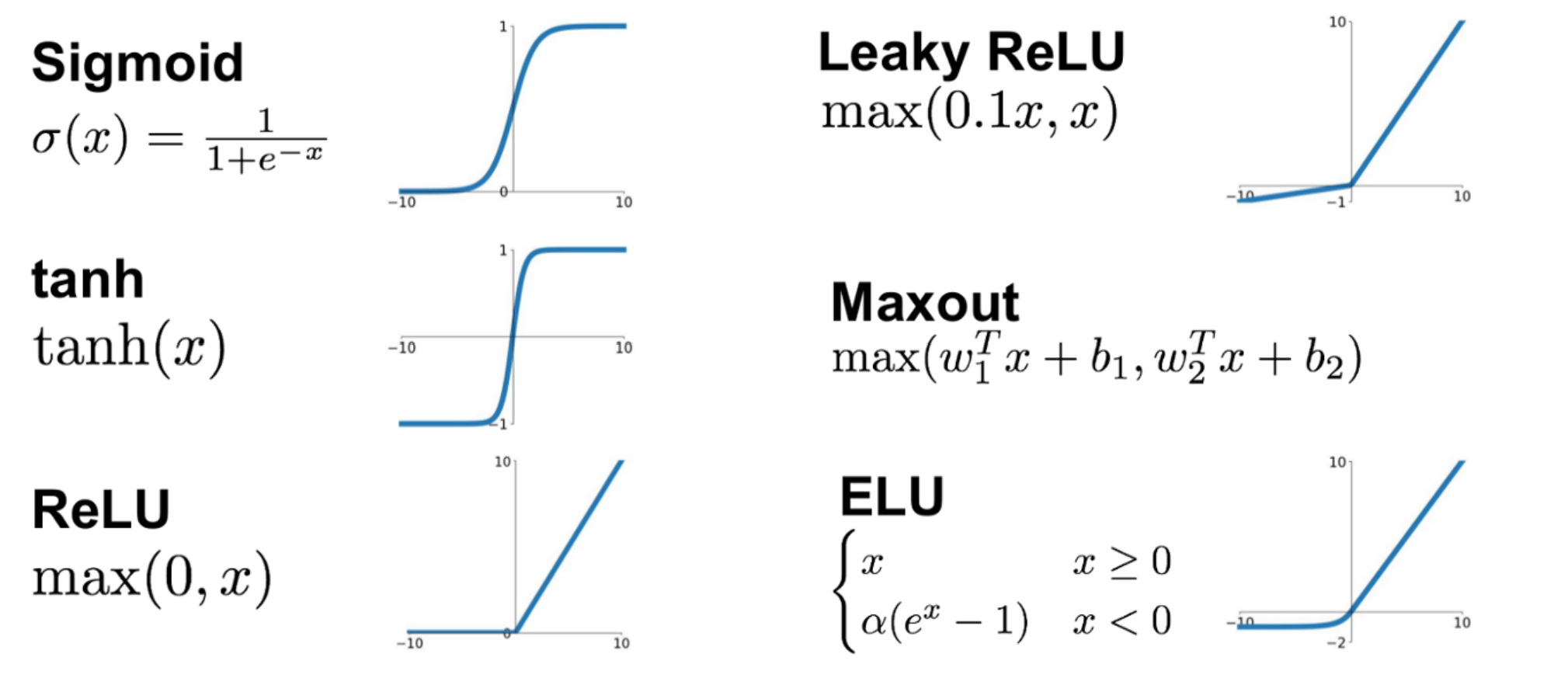}
    \end{center}
    \caption{Activation functions} 
    \label{fig2.1'}
    \end{figure}
    \end{center}
    
    Being known as nonlinear transformation functions, activation layers transform their input values to fall within an acceptable and practical range. There are different types of activation functions (as shown in Fig.\ref{fig2.1'}); however, the rectified linear unit (ReLU) function is the most normally utilized activation function in deep learning. This function keeps the value of non-negative inputs and allocates zeros to negative inputs. That leaves the size of the input volume unchanged. The rectifier function is given below:
    
    \par
    \[
    f(x) =
    \left\{
    \begin{array}{cc}
    x & if \hspace{0.2 cm} x \succ 0\\
    0 & otherwise
    \end{array}
    \right.
     \]
    
    Simplicity and computational efficiency are two main advantages of ReLU. Compared to sigmoid and tanh (the other two commonly used activation functions), the gradient computation for ReLU is much more simple and easy. The gradients are all 1s for non-negative inputs, whereas they are all 0s for negative inputs. Hence, at network training it can remarkably shorten computational time.

\subsection{Pooling Layers}
    
     \begin{center}
    \begin{figure}[htp]
    \begin{center}
     \includegraphics[scale=0.7]{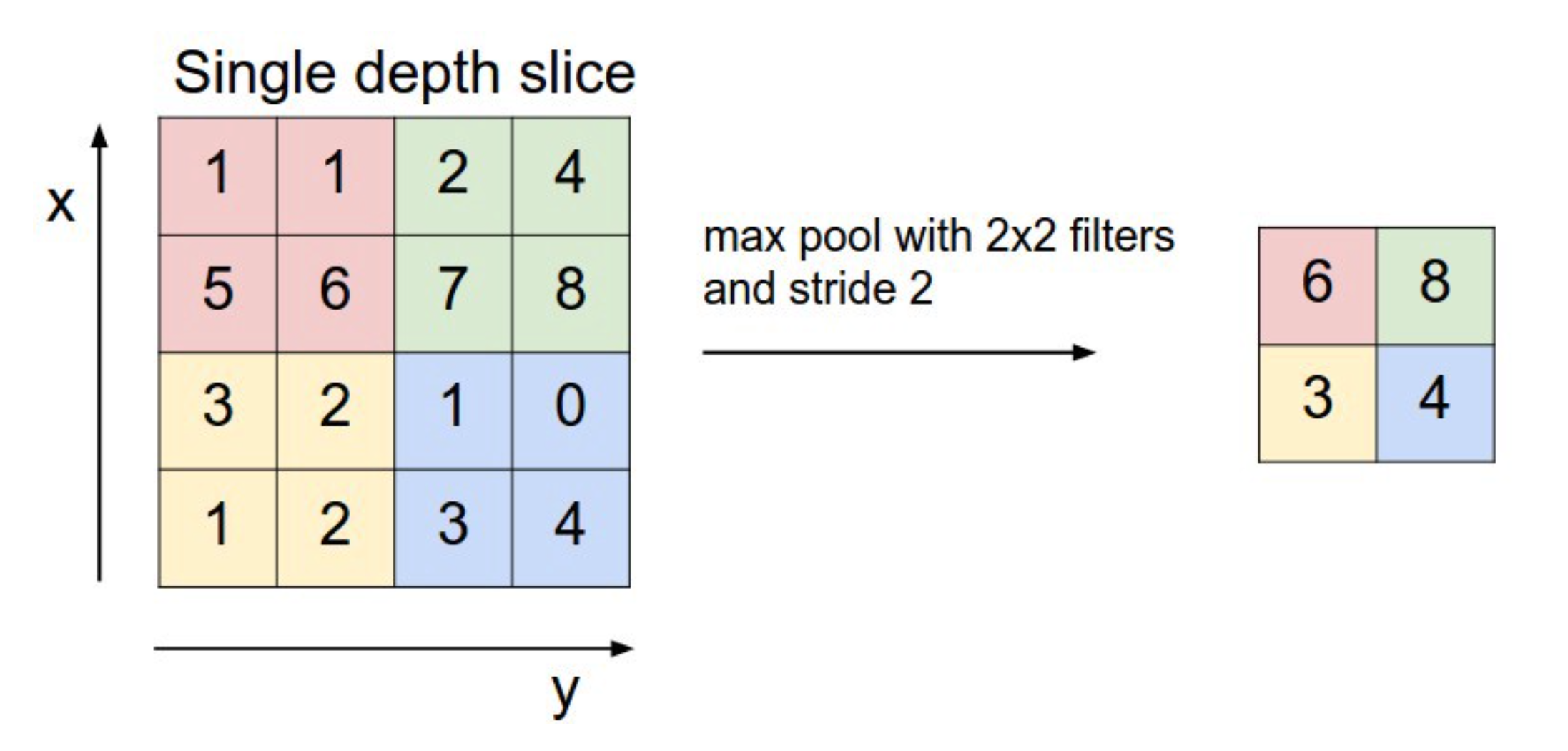}
    \end{center}
    \caption{ Max pooling} 
    \label{fig2.2'}
    \end{figure}
    \end{center}
    
    Pooling layers, also known as downsampling operations, are used to progressively reduce the spatial size of features, the amount of parameters as well as computation in the network, and therefore control overfitting. And, more importantly, they remove/suppress redundant features for improved generalization. Similar to the convolutional layer, the pooling operation sweeps a filter across the entire input, but the difference is that this filter does not have any weights. There are two main types of pooling:
    \begin{itemize}
        \item \textbf{Max pooling}: As the filter moves across the input, it selects the pixel with the maximum value to send to the output array. 
        \item \textbf{Average pooling}: As the filter moves across the input, it computes the average value within the receptive field to send to the output array.
    \end{itemize}
    
    So far the most typically utilized pooling layer has been Max Pooling (as shown in Fig.\ref{fig2.2'}) which partitions the input into a set of non-overlapping regions by a stride of 2 and then outputs the maximum value for each sub-region.
    
\subsection{Fully Connected Layers}
    
\begin{center}
    \begin{figure}[htp]
    \begin{center}
     \includegraphics[scale=0.7]{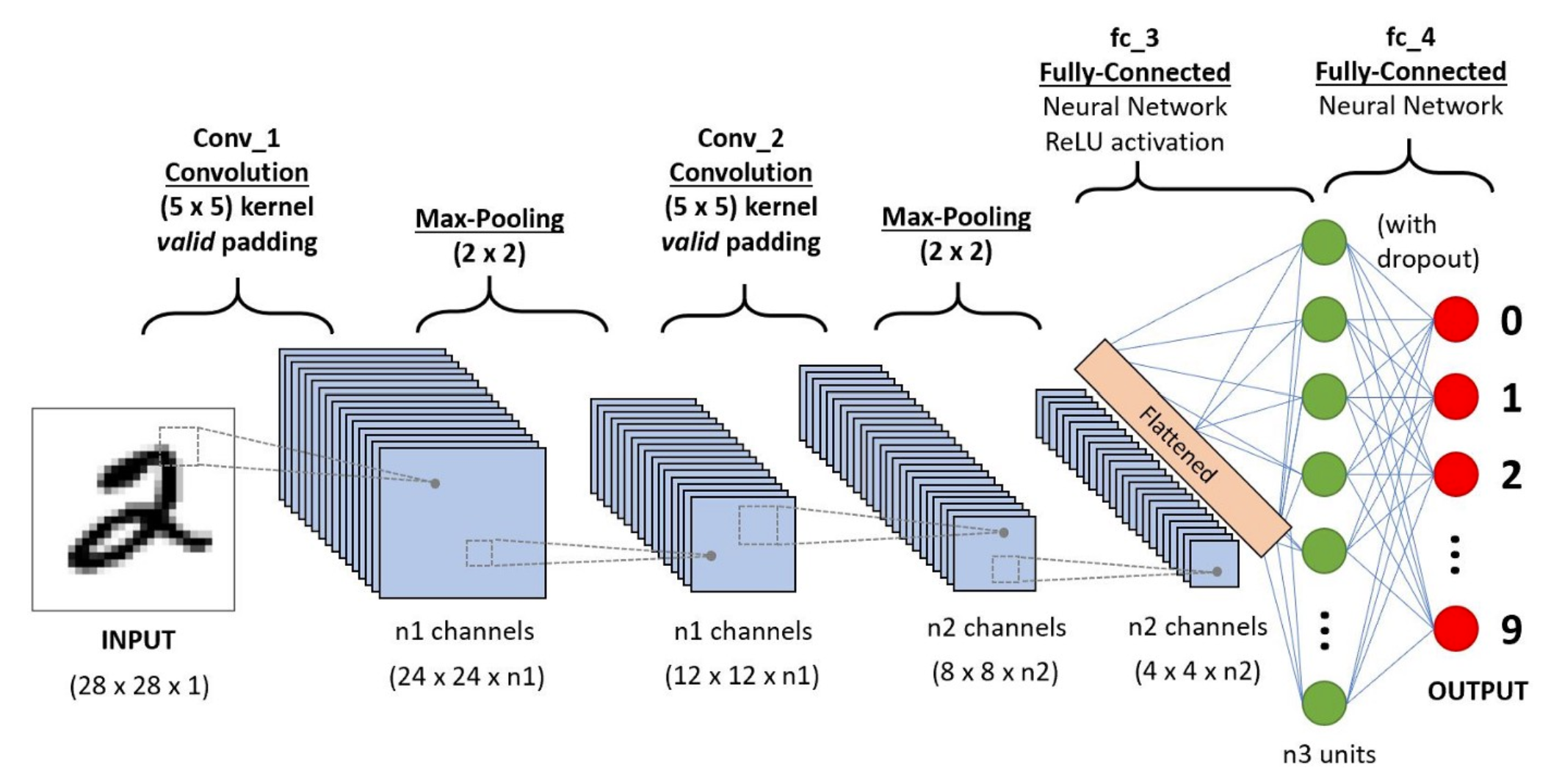}
    \end{center}
    \caption{A Convolutional neural network sequence to classify handwritten digits}
    \label{fig2.3'}
    \end{figure}
    \end{center}

    The name of the full-connected layer (FC layer) aptly describes itself. A fully connected layer consists of a set of neurons where each of them has full connections to its inputs (see Fig.\ref{fig2.3'}, fully connected layers are the last phases for a Convolutional neural network). Given a set of features, activation maps can be computed by performing matrix multiplication plus bias offsets.
    
    This layer performs the task of classification based on the features extracted through the preceding layers and their various filters. Whereas convolutional layers usually follow by ReLu functions, FC layers tend to employ a softmax activation function to classify inputs appropriately, yielding a probability from 0 to 1.
    
\subsection{Normalization Layer}
    
    Outside of the above basic layers, to standardize the statistics of inputs to layers we apply another family of layers called normalization layers which is commonly inserted between a convolution layer and its subsequent activation layer. Because normalization layers can restrict the distributions of inputs to layers in a deep network gently, it will assist the network to yield better gradients for weight update. Consequently, it makes the gradient explosion and vanishing problems less severe during the network optimization~\cite{7298594}. Moreover, training deep neural networks with tens of layers will become more challenging and time-consuming without normalization layers since networks can be very sensitive to the initial random weights and the change in the distribution of network activations during training. There are several normally utilized normalization layers such as batch normalization~\cite{7298594}, layer normalization~\cite{https://doi.org/10.48550/arxiv.1607.06450}, and instance normalization~\cite{https://doi.org/10.48550/arxiv.1607.08022}, which normalize inputs batch-wise, layer-wise, and instance-wise respectively.

\section{Commonly used Deep Neural Networks}

\subsection{Convolutional Neural Networks (CNNs)}

\begin{center}
    \begin{figure}[htp]
    \begin{center}
     \includegraphics[scale=1.6]{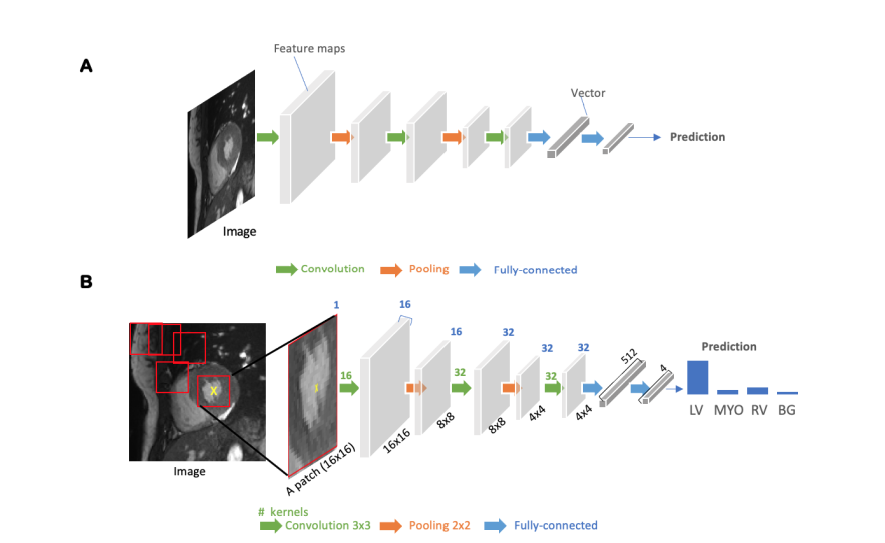}
    \end{center}
    \caption{ Generic architecture of convolutional neural networks (A) and Patch-based segmentation method based on a CNN classifier (B)}
    \label{fig2.1}
    \end{figure}
\end{center}

CNNs are the most popular type of deep neural networks for image analysis, which have been applied to improve the performance of lots of image classification, object detection and segmentation tasks successfully. A typical CNN includes an input layer, an output layer, and a stack of functional layers in between which transform an input into an output in a specific form, e.g., vectors ( shown in Fig.\ref{fig2.1} A). These functional layers are usually convolutional layers, pooling layers, and/or fully connected layers. A convolutional layer CONV$_l$ generally consists of $k_l$ convolution filters/kernels, which is followed by a normalization layer (e.g., layer normalization~\cite{https://doi.org/10.48550/arxiv.1607.06450}), and a nonlinear activation function (e.g., ReLU) to generate $k_l$ feature maps from the input. Following from that, we use the pooling layers to down-sample these feature maps, commonly by a factor of 2 to remove/suppress redundant features, thus improving the statistical efficiency and model generalization. We then utilize fully connected layers to lower the dimension of features from its previous layer and get the most task-relevant features for inference. Finally, the output of the network will be a fix-sized vector where each element can be a probabilistic score for each category (for image classification), a set of values (e.g., the coordinates of a bounding box for object detection and localization), or a real value for a regression task (e.g., the left ventricular volume estimation).

Being the core components of CNNs, each convolutional layer has $k_l$ convolution kernels to extract $k_l$ feature maps. Generally, the size of each kernel $n$ is chosen to be small, e.g., $n = 3$ for a 2D $3\times 3$ kernel to lower the number of parameters\footnote{In a convolution layer $l$ with $k_l$ 2D $n \times n$ convolution kernels and a $l_{in}$-channel input, the number of parameters in a convolutional layer is $k_l \times (n^2 \times l_{in} + 1)$. For example, If we have a convolutional layer with 16 $3 \times 3 $ filters where the input is a $28 \times 28 \times 1$ 2D gray image, the number of parameters that we will have in this layer is $16 \times (3^2 \times 1 + 1) = 160$.}. Consequently, one can increase the receptive field\footnote{The receptive field is the input image area that potentially impacts the activation of a particular convolutional kernel/neuron.} by increasing the number of convolutional layers. For instant, one can utilize three convolutional layers with small $3\times 3$ kernels~\cite{Simonyan:2015} instead of using a convolutional layer with large $7\times 7$ kernels. In this case, the receptive field is the same $(7 \times 7)$ while the number of weights decreases by a factor of $7^2/(3 \times (3^2)) \approx 2$. Widening the receptive field by increasing the number of hidden layers (the depth of convolution neural networks) can result in better model performance, e.g. classification accuracy~\cite{Simonyan:2015} in general.

CNNs used for image classification can also be used for image segmentation applications by adding one more step, which is dividing each image into patches and then training a CNN to predict the class label of the center pixel for every patch (as shown in Fig.\ref{fig2.1} B). The network, in this case has to be deployed for every patch individually even though there is a lot of redundancy due to multiple overlapping patches in the image. As a result of this big downside, CNNs with fully connected layers are commonly applied for object localization, which focusing on estimating the bounding box of the interested object in an image. It is worth noting that this bounding box is then utilized to crop the image in the image pre-processing step to restrict the computational cost for segmentation~\cite{Avendi:2016}. To have a more efficient segmentation, a special type of CNNs called fully convolutional neural network (FCN)/ end-to-end pixel-wise segmentation network is more normally applied. It will described in the next section.

The application of CNN are discribed in Fig.\ref{fig2.1}:
\begin{itemize}
    \item \textbf{ Fig.\ref{fig2.1} A}: a generic architecture of convolutional neural networks. A CNN receives a cardiac MR image as input, learning hierarchical features through a stack of convolutions and pooling operations. Through fully connected layers, these feature maps are then flattened and decreased into a vector. This vector can be in many forms, relying on the specific task. For image classification, it can be probabilities for a set of classes; for object localization, it can be coordinates of a bounding box; for patch-based segmentation, it can be a predicted label for the center pixel of the input; or for regression tasks, it can be a scalar, e.g., left ventricular volume estimation.
    \item \textbf{ Fig.\ref{fig2.1} B}: an example of a patch-based segmentation method based on a CNN classifier. The CNN receives a patch as input and outputs the probabilities for four classes. The prediction for the center pixel (see the yellow cross) in this patch is the class with the highest score. One can finally get a pixel-wise segmentation map for the whole image by repeatedly forwarding patches located at different locations into the CNN for classification. LV: left ventricle cavity; RV: right ventricle cavity; BG: Background; MYO: left ventricular myocardium. The blue number at the top stands for the number of channels of the activation maps. Each convolution kernel is a $3\times3$ kernel (stride=1, padding=1), yielding an output activation map with the same height and width as the input.
\end{itemize} 

\subsection{Fully convolutional neural networks (FCNs)}

\begin{center}
    \begin{figure}[htp]
    \begin{center}
      \includegraphics[scale=1.8]{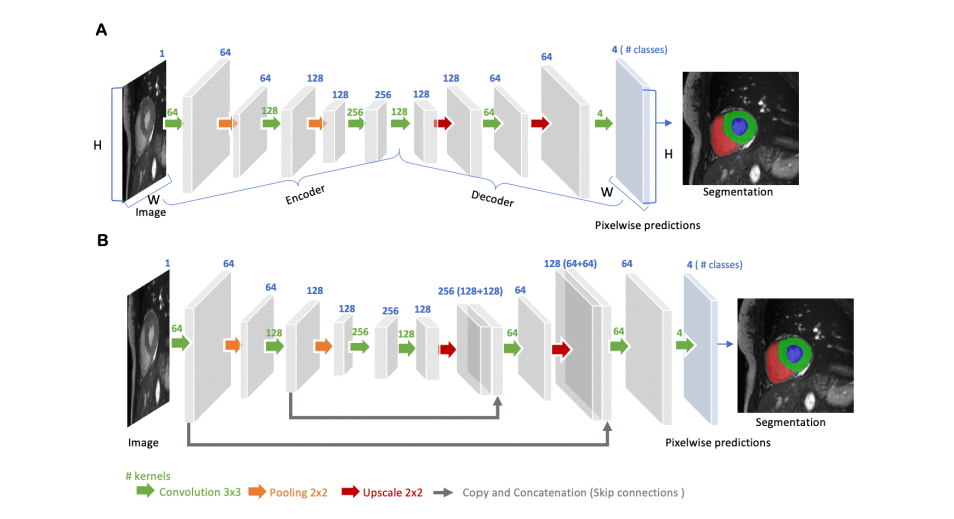}
    \end{center}
    \caption{Architecture of a fully convolutional neural network (FCN) (A) and Architecture of a U-net (B)}
    \label{fig2.2}
    \end{figure}
 \end{center}

 FCNs are special types of CNNs that do not consist of any fully connected layers. The idea of FCN was initially proposed by~\cite{7298965} for image segmentation. As shown in Fig.\ref{fig2.2} A, being designed with an encoder-decoder structure, FCNs can take inputs of arbitrary size and generate an output with the same size. For instance, to perform a segmentation task, the encoder first transforms an input image into a high-level feature representation. Next, the decoder interprets the feature maps and recovers spatial details back into the image space for pixel-wise prediction through a series of upsampling and convolution operations. To accomplish upsampling, one can utilises transposed convolutions (e.g., $3\times3$ transposed convolutional kernels with a stride of 2), or unpooling layers, or upsampling layers to up-scale feature maps (e.g., by a factor of 2). Generally, for segmentation, a FCN outperforms a patch-based CNN as it is trained and applied to the entire images without the need for patch selection~\cite{PMID:27244717}. 

 As displayed in Fig.\ref{fig2.2} A, the simple structure FCNs may have restricted ability to get detailed contextual information in an image for accurate segmentation since it is possible that the pooling layers in the encoder will remove some features. To transmit features from the encoder to the decoder in order to improve the segmentation precision, there are several successful variants of FCNs that have been recommended. Among those, the U-net is the most famous and most intensive variant of FCNs for biomedical image segmentation~\cite{RFB15a}. Base on the vanilla FCN, a U-net (see Fig.\ref{fig2.2} B) utilizes skip connections between the encoder and decoder to retrieve spatial context loss in the down-sampling path, producing more accurate segmentation. Recently, the U-net or its 3D variants, the 3D U-net~\cite{10.1007/978-3-319-46723-8_49} and the 3D V-net~\cite{7785132} has been employed for several ultramodern medical image segmentation methods as their backbone networks to yield expecting segmentation precision~\cite{163c9ff9045b4801a19765176aa1bff4,10.1007/978-3-319-75541-0_13,Xia2018Automatic3A}.

The frameworks of FCNs are discribed in Fig.\ref{fig2.2}:
\begin{itemize}
    \item \textbf{ Fig.\ref{fig2.2} A}: An architecture of a fully convolutional neural network (FCN). The encoder first gets the whole image as input to learn image features. Next, the decoder gradually recovers the spatial dimension by a series of upscaling layers (e.g., transposed convolution layers, unpooling layers) and then yields pixel-wise probabilistic maps to predict regions of the left ventricle cavity (blue region), the left ventricular myocardium (green region) and the right ventricle cavity (red region). By allocating each pixel with the class of the highest probability, one acquires the final segmentation map. One use case of this FCN-based cardiac segmentation can be found in~\cite{https://doi.org/10.48550/arxiv.1604.00494}.
    \item \textbf{ Fig.\ref{fig2.2} B}: An architecture of a U-net. U-net adds ‘skip connections’ (gray arrows) on the basis of FCN to assemble feature maps from coarse to fine through concatenation and convolution operations. The number of downsampling and upsampling blocks in the diagram is decreased for simplicity. The detailed information can be found at the original paper~\cite{RFB15a}.
\end{itemize}

\subsection{Recurrent Neural Networks (RNNs)}

\begin{center}
    \begin{figure}[htp]
    \begin{center}
     \includegraphics[scale=1.4]{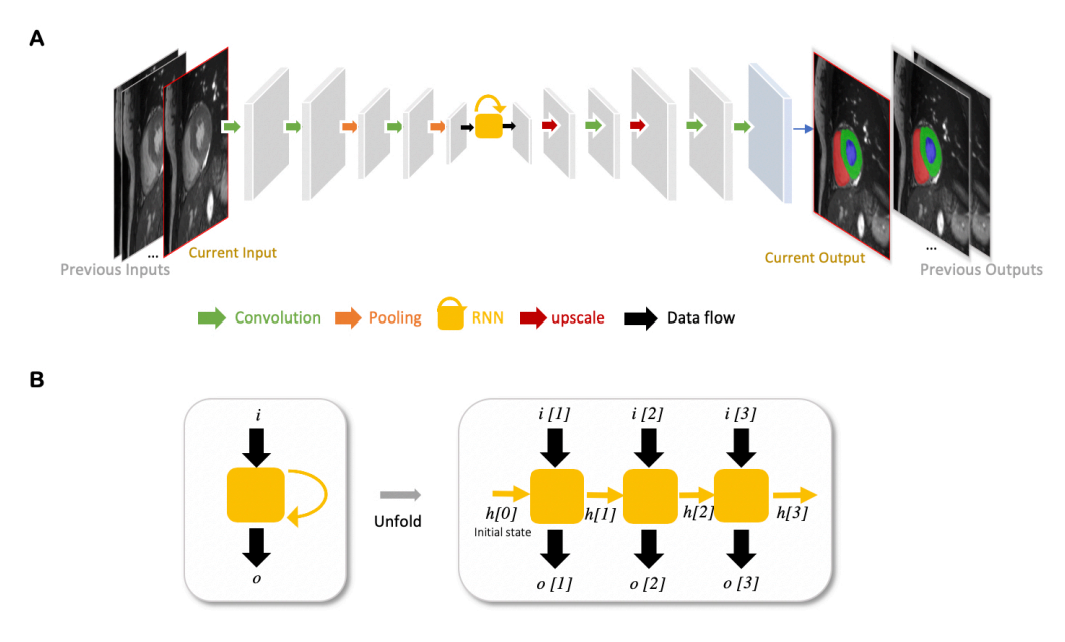}
    \end{center}
    \caption{Example of an FCN with an RNN for cardiac image segmentation (A) and Unfolded schema of the RNN module for visualizing the inner process when the input is a sequence of three images (B)}
    \label{fig2.3}
    \end{figure}
\end{center}

Another common type of deep neural networks are Recurrent neural networks (RNNs). They are adopted for sequential data, such as cine magnetic resonance imaging (MRI) and ultrasound image sequences. An RNN can generate present decision by 'memorizing' the past and using the knowledge which is learned from this past (see Fig.\ref{fig2.3} A and B). For instance, given a sequence of images, an RNN will use the first image as input, get the information to yield a prediction, and remember this information that is then employed to predict the next image. Long-short term memory (LSTM)~\cite{10.1162/neco.1997.9.8.1735} and gated recurrent unit (GRU)~\cite{Cho2014LearningPR} are the two most commonly employed architectures in the family of RNNs since they have the capacity for modeling long-term memory. In practice, one can combine an RNN with a 2D FCN for cardiac segmentation so that the combined network is able to manage catching information from adjacent slices to advance the inter-slice coherence of segmentation outcomes~\cite{10.1007/978-3-319-52280-7_8}.

The application of RNNs are displayed in Fig.\ref{fig2.3}:
\begin{itemize}
    \item \textbf{ Fig.\ref{fig2.3} A}: an example of an FCN with an RNN for cardiac image segmentation. The RNN module is deputized by the yellow block with a curved arrow, which utilizes the knowledge learned from the past to yield the present decision. Here, the network is adopted to segment cardiac ventricles from a stack of 2D cardiac MR slices, which enables the propagation of contextual information from adjacent slices for better inter-slice coherence~\cite{10.1007/978-3-319-52280-7_8}. Sequential data such as cine MR images and ultrasound movies also apply this type of RNN to learn temporal coherence.
    \item \textbf{ Fig.\ref{fig2.3} B}: visualizing the inner process for an unfolded schema of the RNN module when the input is a sequence of three images. Each time, this RNN module gets an input $i[t]$ at time step t, and generates an output $o[t]$. It considers not only the input information but also the hidden state ('memory') $h[t-1]$ from the preceding time step $t - 1$.
\end{itemize}

\subsection{Autoencoders (AE)}

\begin{center}
    \begin{figure}[htp]
    \begin{center}  
     \includegraphics[scale=1.4]{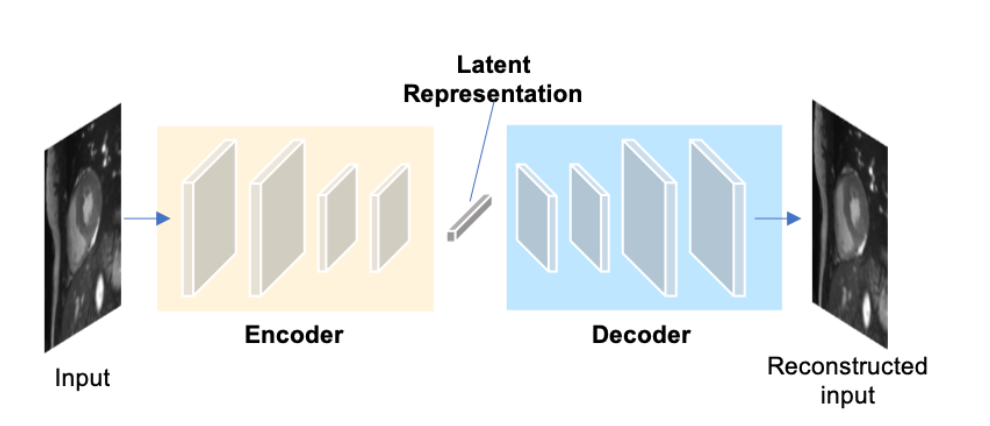}
    \end{center}
    \caption{Generic architecture of an autoencoder}
    \label{fig2.4}
    \end{figure}
\end{center}

Another type of neural network is Autoencoder (AE), which consists of an encoder network and a decoder network to reconstruct the input (shown in Fig.\ref{fig2.4}). AEs are capable of learning compact latent representations from data without supervision. As those learned representations consist of commonly useful information in the primary data, these neuron networks have been adopted to produce general semantic features or shape information from input images or labels. Those useful features are then employed to conduct the medical image segmentation~\cite{10.1007/978-3-030-00928-1_30,10.1007/978-3-319-46726-9_29,10.1007/978-3-030-32245-8_62}.

As shown in Fig.\ref{fig2.4}, an autoencoder uses an encoder-decoder framework. The encoder first transforms the input data to a low-dimensional latent representation. The decoder then expounds the code and rebuilds the input. One has found that the learned latent representation is effective for cardiac image segmentation~\cite{10.1007/978-3-030-00928-1_30,10635_150626}, cardiac shape modeling~\cite{10.1007/978-3-030-00934-2_52} and cardiac segmentation correction~\cite{10.1007/978-3-030-32245-8_70}.

\subsection{Generative Adversarial Networks (GANs)}

\begin{center}
    \begin{figure}[htp]
    \begin{center}
     \includegraphics[scale=1.4]{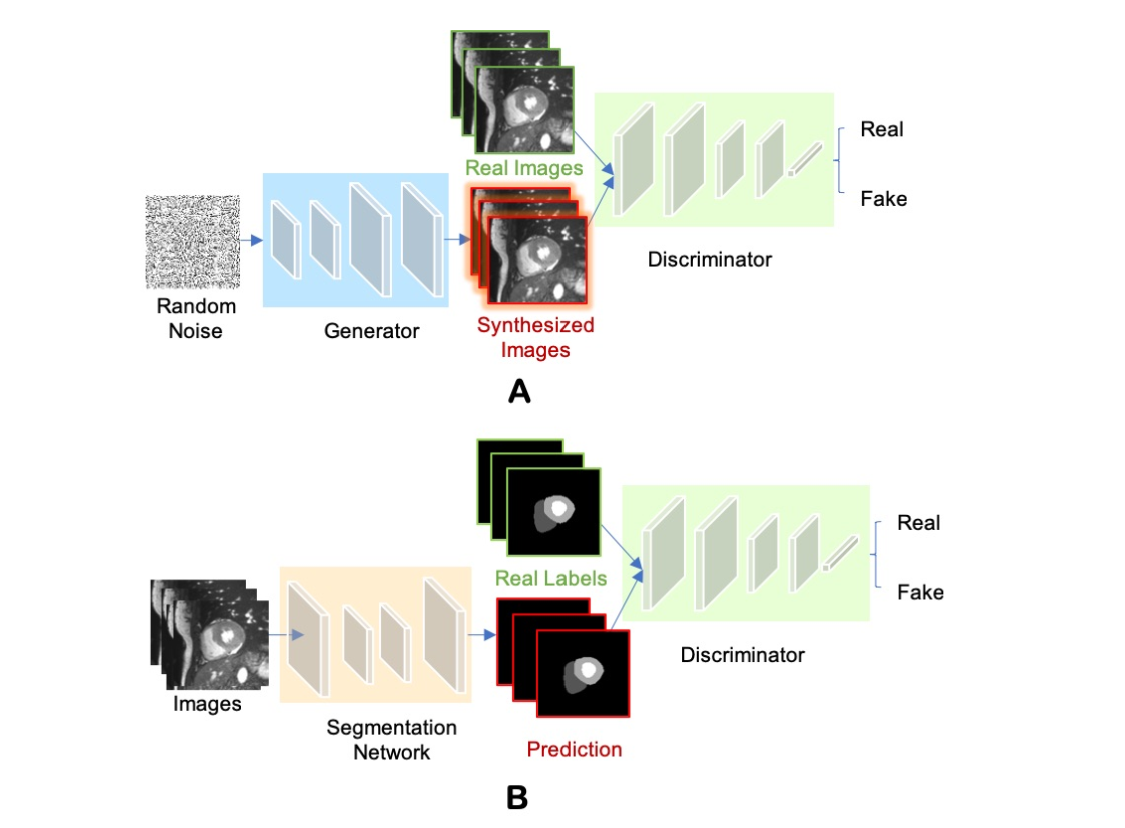}
    \end{center}
    \caption{Overview of GAN for image synthesis (A) and Overview of adversarial training for image segmentation (B)}
    \label{fig2.5}
    \end{figure}
\end{center}

Being recommended by~\cite{NIPS2014_5ca3e9b1} for image synthesis from noise, Generative Adversarial Networks are a type of generative models that learn to model the data distribution of real data and consequently are capable of producing new image examples. A GAN includes two main components: a generator network and a discriminator network which are trained to compete against each other during training (see Fig.\ref{fig2.5} A). The generator yields fake images to fool the discriminator intentionally, whereas the discriminator tries to differentiate fake images from real ones. As the two models are both set to win this competition, this type of training is called 'adversarial training', which can also be employed for training a segmentation network. For the segmentation task, as shown in Fig.\ref{fig2.5} B, the generator is replaced by a segmentation network and the discriminator is expected to differentiate the ground truth ones (the target segmentation maps) from the produced segmentation maps. Consequently, one can utilizes the segmentation network to yield more plausible segmentation maps anatomically~\cite{luc:hal-01398049,Savioli2018AGA}.

\section{Training Deep Neural Networks}

\subsection{Overview}
Neural networks must be trained before they are capable of realizing inference. In a standard training process, the network parameters will be adjusted to reduce the difference between the actual outputs from the network and the desired outputs function as small as possible. Consequently, we need a dataset that consists of paired images and labels for both training and testing, an optimizer (e.g., stochastic gradient descent (SGD)~\cite{10.1214/aoms/1177729586}, or adaptive moment estimation (Adam)~\cite{Kingma2015AdamAM}) and a loss function (or sometimes also referred to as the cost function or the objective) to update all parameters. This loss function represents the error of the model prediction in every iteration, which also supplies signals for the optimizer to adjust the model parameters through back-propagation~\cite{Rumelhart1986LearningRB}. And during training we aim to choose best values of the model parameters which minimize the loss function.

It can be formulated as a minimization problem in Mathematics. Assume we have $f$ as a neural network with a set of learnable parameters $\theta$ (e.g., weights $w$ and biases $b$ in each convolutional layer), we need to find optimal $\theta^*$ as below:

\begin{equation}
    \theta^* = \mathrm{arg}\min_{\theta} \mathcal{L}_{exp} = \mathrm{arg}\min_{\theta} \mathbb{E}_{(x,y) \sim P(X,Y)} \mathcal{L} [\textbf{y},f(\textbf{x},\theta)],
\end{equation}
where 
\begin{itemize}
    \item $x \in \textbf{X}$ is an input image and $y \in \textbf{Y}$ is a corresponding target label;
    \item $\mathbb{E}_{(x,y) \sim P(X,Y)} \mathcal{L} [\textbf{y},f(\textbf{x},\theta)]$ is the expected loss over the joint distribution $P(X, Y)$.
\end{itemize}

In practice, we will find $\theta$ that minimizes the empirical loss/risk which computed only on a given dataset (e.g., training set) $\mathcal{D}_{tr}$ because we don't know the joint distribution of $P(X, Y)$. Hence, we instead will find an proximate solution $\hat{\theta} $ of $\theta^*$:
\begin{equation}
    \hat{\theta} = \mathrm{arg}\min_{\theta} \mathcal{L}_{emp} = \mathrm{arg} \min_{\theta}\mathbb{E}_{(x,y)\sim \mathcal{D}_{tr}}\mathcal{L}[\textbf{y},f(\textbf{x},\theta)].
\end{equation}

This approach is also called empirical risk minimization (ERM)~\cite{788640} in the statistical learning theory. This theory suggests that we should select a hypothesis space which minimizes the empirical risk for the learning algorithm.

\subsection{Back-propagation}

In the network learning process, the backpropagation (BP) algorithm~\cite{Rumelhart1986LearningRB} is a central concept, which minimizes the training loss by updating the model parameters during network training. BP calculates the gradients layer by layer at a high level from the very last layer to the earlier layers. One then applies gradient descent to adjust the associated weights in the direction that minimizes the difference/error between the desired outputs functions/ground true labels and the outputs from the network/predictions. There are four steps in the learning process particularly:
\begin{itemize}
    \item Step 1: pass the input data \textbf{$x$} to the network $f(\cdot;\theta)$ which parameterised by $\theta$, and then calculate predictions $f(\textbf{x};\theta)$;
    \item Step 2: calculate the loss $\mathcal{L}[\textbf{y}, f(\textbf{x}; \theta)]$ between the desired outputs $\textbf{y}$ and the actual network outputs $f(\textbf{x}; \theta)$;
    \item Step 3: backpropagate the loss from the final layers to preceding layers by applying chain rule repeatedly to calculating the gradients of the loss/errors with respect to the network parameters $\Delta_{\theta}\mathcal{L}$ layer by layer;
    \item Step 4: select a gradient descent algorithm, e.g., SGD to adjust those parameters $\theta: \theta \leftarrow \theta - \lambda\Delta_{\theta}\mathcal{L}$ where $\lambda$ is the step size.
\end{itemize}

\subsection{Loss Functions}

In practice, there are various types of problems that we want to solve (e.g., regression, classification, etc.). One can choose from several different widely-used loss functions $\mathcal{L}$ to achieve the expected results. These loss functions are:

\textbf{Mean squared error (MSE)}

MSE is the simplest loss function for regression, which is the task of predicting real-valued quantities, such as the price of houses or the length of something in an image. For this task, it is common to calculate the loss between the true value and the predicted quantity which is defined as below:

\begin{equation}
    \mathcal{L}_{\mathrm{MSE}} = \frac{1}{n}\sum_{i=1}^n(\mathbf{y_i}-\mathbf{\hat{y}_i})^2,
\end{equation}
where 
\begin{itemize}
    \item \textbf{$y$} is the vector of target/desired values and $\mathbf{\hat{y}} = f(\mathbf{x}; \theta)$ is the vector of the predicted values;
    \item The subscript $i$ describes the $i$-th element in the corresponding vector;
    \item $n$ is the length of each vector.
\end{itemize}

\textbf{Cross-entropy}

For both image classification and segmentation tasks, the most commonly-used loss function is Cross-entropy where the model yields the probability for each class instead of class labels\footnote{The predicted segmentation map for each image is achieved by allocating each pixel with the class of the highest probability:$\mathbf{\hat{y}_i}=\mathrm{arg}\max_c\mathbf{p}_i^c$ at inference time.}. Specifically, for segmentation, the Cross-entropy loss sums up the pixel-wise probability errors between the predicted probabilistic output from the network after softmax $\mathbf{p}^{(c)} = \frac{e^{f(x,\theta)^{(c)}}}{\sum_{d=1}^C e^{f(x,\theta)^{(d)}}}$ and its corresponding target one-hot segmentation map $\mathbf{y}^{(c)}$ for each class $c$:
\begin{equation}
    \mathcal{L}_{CE(segmentation)}=-\frac{1}{n}\sum_{i=1}^n\sum_{c=1}^C\mathbf{y}_i^{(c)}\log(\mathbf{p}_i^{(c)}),
\end{equation}
where $n$ is the number of pixels and $C$ is the number of all classes in the corresponding image. On the other hand, one can obtain the loss for image-level classification tasks by simply removing the pixel-wise summation: 
\begin{equation}
    \mathcal{L}_{CE(classification)}=-\sum_{c=1}^C\mathbf{y}^{(c)}\log(\mathbf{p}^{(c)})
\end{equation}

\textbf{Soft-Dice }

Soft-Dice is another loss function, which is particularly formulated for object segmentation~\cite{7785132}. This loss function penalizes the mismatch between a target map and its predicted segmentation map at pixel-level:
\begin{equation}
    \mathcal{L}_{Dice}=1-\frac{2\sum_{i=1}^n\sum_{c=1}^C\mathbf{y_i^{(c)}\mathbf{p_i^{(c)}}}}{\sum_{i=1}^n\sum_{c=1}^C(\mathbf{y_i^{(c)}}+\mathbf{p}_i^{(c)})}
\end{equation}

\textbf{Several variants of the cross-entropy  and soft-Dice}

Besides the above typical loss functions, there are different versions of the cross-entropy and soft-Dice loss such as the weighted cross-entropy loss~\cite{10.1007/978-3-319-75541-0_17,10.1007/978-3-319-75541-0_12,Yang2017ClassBalancedDN} and weighted soft-Dice loss~\cite{KHENED201921}. They are mainly employed in medical image segmentation tasks to solve potential class imbalance issue where the loss term is weighted for rare classes or small objects. For instant, the weighted cross-entropy loss is designed as:
\begin{equation}
    \mathcal{L}_{Weighted CE}=-\frac{1}{n}\sum_{i=1}^n\sum_{c=1}^Cw^{(c)}\mathbf{y}_i^{(c)}\log(\mathbf{p}_i^{(c)}),
\end{equation}
where $w_{(c)}$ is a scalar, determining the weight for the loss term associated with the class $c$. In practice, $w_{(c)}$ for a majority class is set to a lower value than the one for the rare class.

\chapter{Related Work}

In this chapter, we review several recent developments of deep learning-based applications for spinal imaging, which is applied widely in the clinical assessment and treatment planning of Adolescent idiopathic scoliosis (AIS). Spine imaging is a non-invasive imaging technique that can visualize the spinal structure. So far, Spine imaging has utilized different imaging methods to estimate the landmarks, segment vertebrae and predict cobb angles. Subsequently, Spine imaging is an essential tool for assessing spinal pathologies.

In the next parts, we briefly describe the latest evolution of spine imaging, with a particular concentration on automatic spine curvature estimation, where the deep learning techniques have been robustly employed in.

\section{Deep Learning for Spinal Imaging}

\subsection{Detection of Vertebral Fractures}

Osteoporosis is one of the main reasons of inducing fractures. Vertebral compression fractures are early signs of the disease, and they are seen as risk predictors for secondary osteoporotic fractures. Therefore, it is important to detect vertebral fractures in CT images. In 2019, J. Nicolaes et al. introduced a very famous method named 'Detection of Vertebral Fractures in CT Using 3D Convolutional Neural Networks'~\cite{10.1007/978-3-030-39752-4_1}, which screens spine-containing CT images opportunistically to detect the existence of these vertebral fractures. While previous methods inspired by radiology practice are built up of 2D and 2.5D features, the proposed method is the first one that employs learned 3D feature maps to automatically detect vertebral fractures in CT. This method localizes exactly these fractures so that radiologists can expound its results. 

A voxel-classification 3D Convolutional Neural Network (CNN) was trained with a training database of 90 cases which has been semi-automatically produced adopting radiologist readings available in clinical practice. Their 3D method outperforms other existing methods with an Area Under the Curve (AUC) of $95\%$ for patient-level fracture detection and an AUC of $93\%$ for vertebra-level fracture detection in experiment.

\textbf{Methods}

\begin{center}
    \begin{figure}[htp]
    \begin{center}
     \includegraphics[scale=1.2]{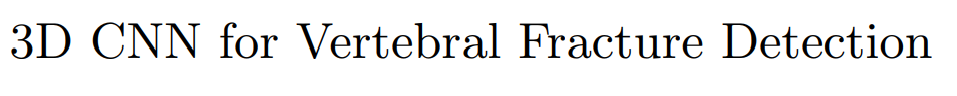}
    \end{center}
    \label{fig3.1'}
    \end{figure}
\end{center}

Their approach requires a label for every voxel in each training image. That is to use a 3D CNN to predict a class probability for every voxel. However, the main task is detection and not segmentation. Consequently, correctly predicting only a sufficient amount of voxels around the vertebra centroid is needed to detect normal or fractured vertebrae in an image. In particular, a radiologist initially created a text file with annotations for every vertebra present in each image. These labels were then enriched with 3D centroid coordinates by manually localizing every vertebra centroid in the images. Lastly, they expended the method described by Glocker et al.~\cite{10.1007/978-3-642-40763-5_33} to automatically produce 3D label images from those sparse annotations. The more details are presented as below:

\begin{itemize}
    \item The Vertebral Fractures detection task is conducted adopting the voxel classification model (as displayed in Fig.\ref{fig3.1'}) that includes an 11-layer dual pathway architecture consisting of 230 K parameters. This CNN contains 8 convolution layers each of which has filters of size $3^3$ yielding an effective receptive field of $17^3$ in the normal pathway and $51^3$ in the subsampled pathway (subsampling factor 3). 
    \item The voxel classifier maps an input image into a prediction image that consists of a probability $p(f|\textbf{x}), f \in F = \left\{background, normal, fracture\right\}$ for every voxel $\textbf{x}$ in the image. This information is then aggregated to patient-level (detecting whether a fracture is present in the patient image) or vertebra-level (detecting whether a fracture is present for every vertebra visible in the patient image).
\end{itemize}

\begin{center}
    \begin{figure}[htp]
    \begin{center}
     \includegraphics[scale=0.9]{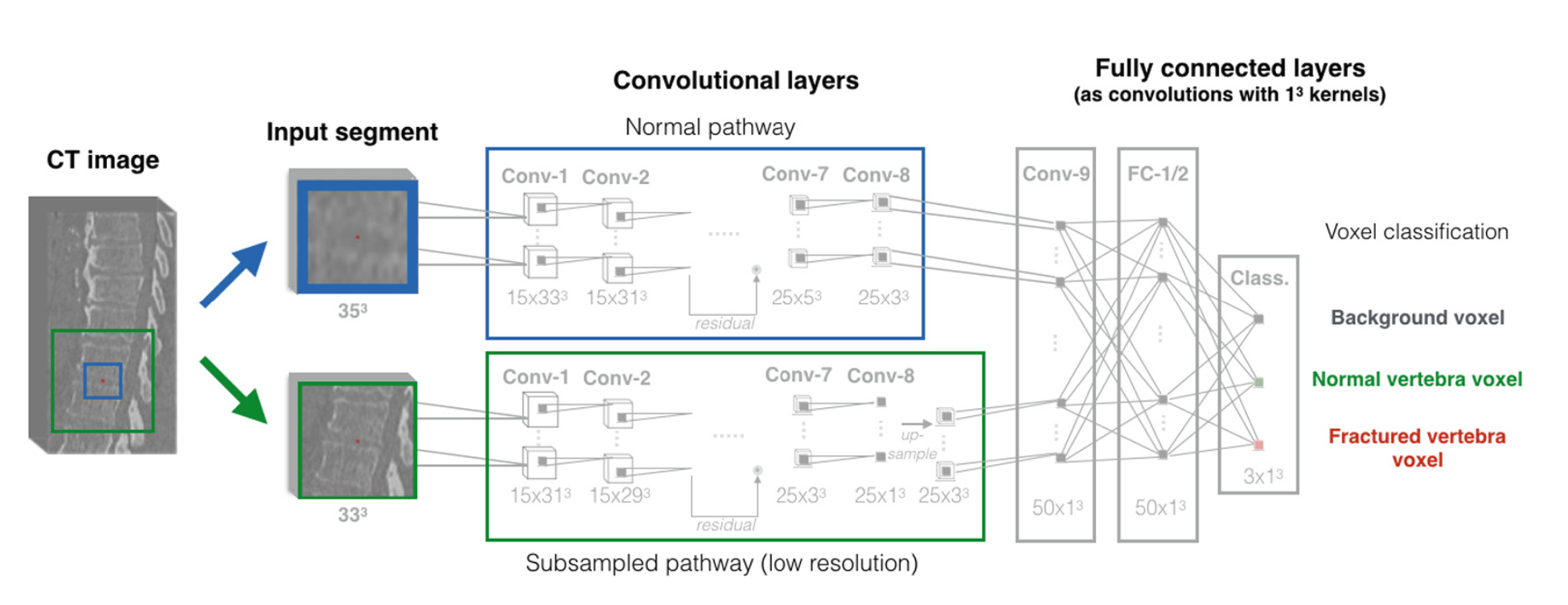}
    \end{center}
    \caption{A 3D CNN 11 - layer dual pathway architecture}
    \label{fig3.1}
    \end{figure}
\end{center}

\textbf{Results}

Their 3D method achieves an Area Under the Curve (AUC) of $0.95 \pm 0.02$ for patient-level fracture detection and an AUC of $0.93 \pm 0.01$ for vertebra-level fracture detection in a five-fold cross-validation experiment, which are comparable to other results reported by other researchers previously. 

\subsection{Vertebral Labelling in Radiographs}

In the treatment planning of scoliosis and degenerative disorders, analyzing spinal shape is significant. In order to achieve this task successfully, it is essential to localize and label vertebrae especially in spinal radiographs. Nevertheless, tissue overlaying and size of spinal radiographs have become challenging problems for vertebrae localization and labeling. Fortunately, A. Bayat et al. introduced a powerful and practical approach named 'Learning a Coordinate Corrector to Enforce Spinal Shape'~\cite{10.1007/978-3-030-39752-4_4}. In this approach, the proposed model has a holistic view of the input image to detect landmark in large and noisy images regardless of its size and employs it on spinal radiographs. According to this model, the labels and locations of vertebrae is predicted in a two-stage model: 
\begin{itemize}
    \item \textbf{stage 1:} the rough location of landmarks is estimated by predicting 2D Gaussians using a fully convolutional network (FCN). Also, the 2D Gaussian image is compared to the ground truth 2D Gaussians at this level;
    \item \textbf{stage 2:} the residual corrector (RC) component is adopted to convert the Gaussians predicted in the previous step to coordinate format and then correct the coordinates.
\end{itemize}

The RC component can be merged to the deep neural network, and trained end-to-end with other sub-networks because of its differentiable functionality. This new approach successfully yields good outcomes that are comparable to all previous state-of-the-art models. 

\textbf{Methodology}

As shown in Fig.\ref{fig3.4}, the model is composed of two sub-networks: U-net as a FCN, and the RC component, which is highlighted with a dashed-line box.

\begin{itemize}
    \item \textbf{In the U-net/FCN:} it produces predicting 2D Gaussians. At this level, The 2D Gaussian image is compared to ground-truth Gaussian images $ Y \in R^{(h\times w\times25)}$, which is a 25-channeled, 2D image with each channel corresponding to each of the 24 vertebrae (C1 to L5), and one for the background; $h$ and $w$ are respectively height and width of the input image. 
    \item \textbf{In the RC component:} first the centroid coordinates are captured adopting softargmax method~\cite{10.1007/978-3-319-46466-4_28,10.1007/s10791-009-9110-3}. Here, the dimensionality of the input data is reduced from $h \times w$ to $2 \times 24$, where 24 is the number of vertebrae. For each vertebra, 2 parameters are estimated as 2D coordinates. Particularly, the dimensionality of data can be reduced from input images with variable size to a fixed size of $2 \times 24$ as long as the input image contains the whole spine. After that, a residual block of fully-connected layers are used to correct the location of the estimated coordinates and increase the receptive field of the model.

\end{itemize}

\begin{center}
    \begin{figure}[htp]
    \begin{center}
     \includegraphics[scale=1.2]{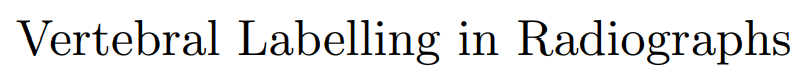}
    \end{center}
    \label{fig3.4'}
    \end{figure}
\end{center}

\begin{center}
    \begin{figure}[htp]
    \begin{center}
     \includegraphics[scale=0.9]{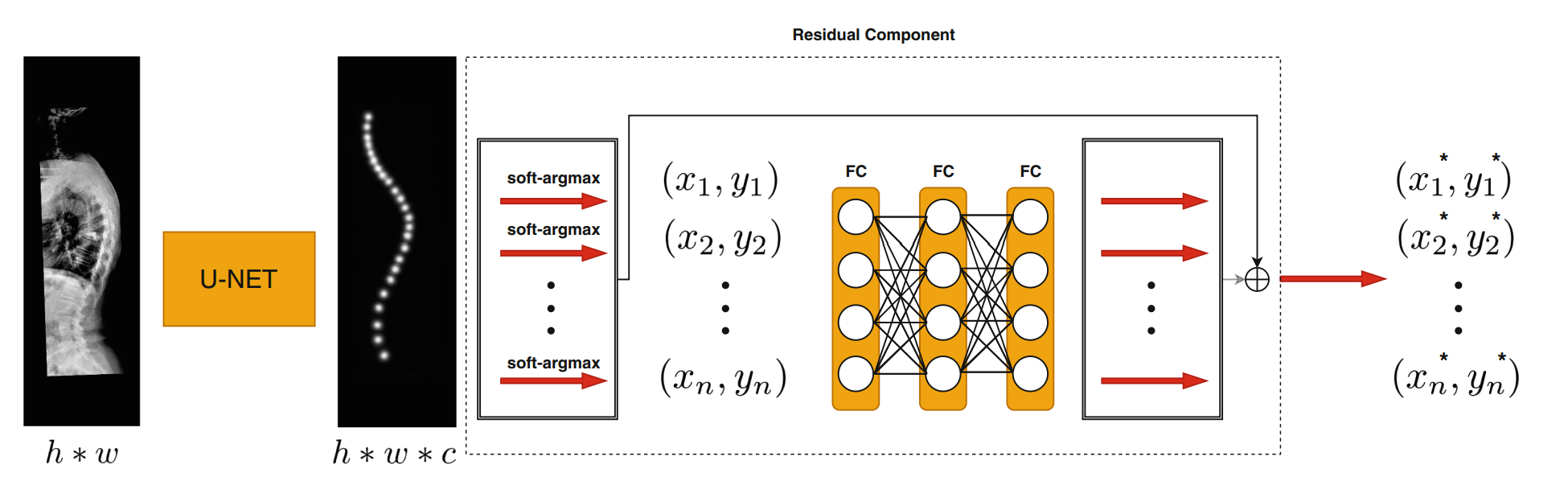}
    \end{center}
    \caption{Overview of the model}
    \label{fig3.4}
    \end{figure}
\end{center}

Since soft-argmax is differentiable, the fully connected layers in RC component sub-network can be trained along with the FCN end-to-end and variable size input images. These fully connected layers give a holistic view of the input image to the network and result in the correct order of label and morphological consistency of the spine.

\textbf{Results}

In practice, the proposed model accomplished respectively identification rates of $85.32\%$ and $52.28\%$ for sagittal and coronal views and localization distance of 4.57 mm and 5.33 mm in sagittal and coronal views radiographs.

\subsection{Segmentation of nerve, Bone, and Disc from 3D spinal images}

Although deep learning has become a promising technology to accomplish the automatic segmentation of medical images, in clinical practice, there are still not many labeled data available for the application of common deep networks. Consequently, it is necessary to grow a deep learning network that can be applied to a few labeled data and analyse its outputs accuracy. For instant, to support preoperative assessment of spinal surgery, there is a real requirement to design a network that is capable of segmenting precisely lumbosacral structures on thin-layer computed tomography (CT) in which labeling those 3D medical data is really time-consuming and costly in reality. 

To address this issue, H. Liu et al developed Semi-cGAN and fewshot-GAN that can automatically segment nerve, bone, and disc from 3D images with remarkable accuracy. Their work is named 'Semi-supervised Semantic Segmentation of Multiple Lumbosacral Structures on CT'~\cite{10.1007/978-3-030-39752-4_5}. To assess the segmentation precision of lumbosacral structures, they employ Dice Score and average symmetric surface distance (ASD) to estimate the segmentation performance of lumbosacral structures. Besides, to test the generalization ability of the trained semi-cGAN and fewshot-GAN, they use another dataset from SpineWeb. Outcomes of the study show that the performance of the two robust models slightly outperform 3D U-Net for automatic segmenting lumbosacral structures on thin-layer CT with fewer labeled data.

\begin{center}
    \begin{figure}[htp]
    \begin{center}
     \includegraphics[scale=0.9]{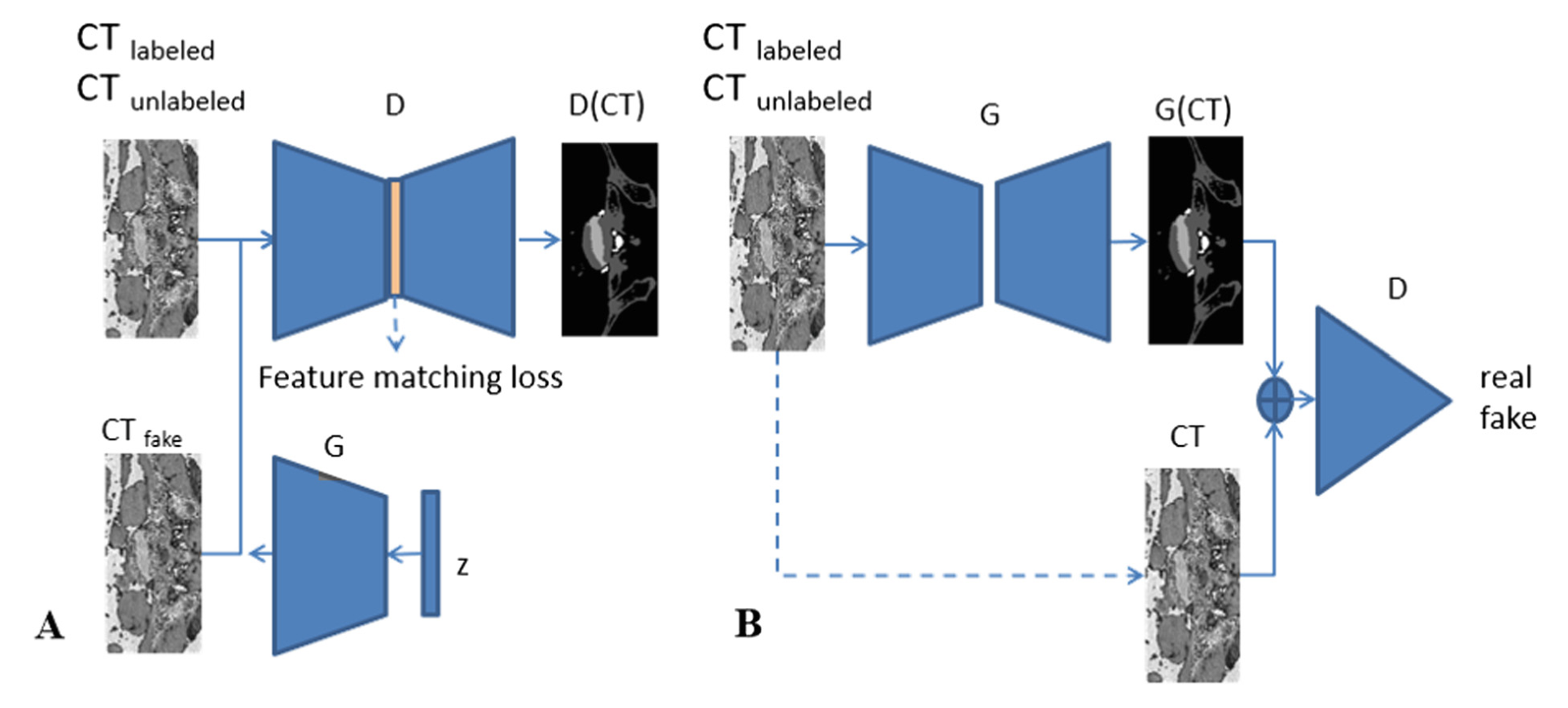}
    \end{center}
    \caption{Schematic drawing of two network architecture. A: Fewshot-GAN network architecture; B: Semi-cGAN network architecture}
    \label{fig3.5}
    \end{figure}
\end{center}

\textbf{Methodology}

\textbf{Fewshot-GAN network architecture:} generator G of fewshot-GAN is a volume generator introduced by Wu et al.~\cite{10.5555/3157096.3157106} (see Fig. \ref{fig3.5} A). Its input is a noise vector z and its output is a fake patch. This fake patch then concatenates to a labeled patch and an unlabeled patch to become input for Discriminator D, which is a modified 3D U-Net with leaky ReLUs, weight-normalization (instead of batch normalization) and average pooling (rather than Max pooling). Discriminator D works as a segmentation network to produce corresponding masks. %Feature matching loss is used to match features of generated images and features in an intermediate layer of discriminator.

\textbf{Semi-cGAN network architecture:} generator G of semi-cGAN is a 3D U-Net that serves as a segmentation network (see Fig.\ref{fig3.5} B). CT patch x (labeled CT or unlabeled CT) and the corresponding masks of automatic segmentation G(x) are respectively the input and output of generator G. G(x) consists of four channels (4 classes: nerve, bone, disc, background), which is the same to manually labeled masks y (unlabeled CT has no corresponding masks y). After that, G(x) or y associates with x to yield a pair $(x, G(x))$ or $(x, y)$ and becomes input to Discriminator D, which is a 3D convolutional neural network (3D-CNN). The last two layers of D are a fully connected layer, and a layer that produces the probability of $(x, G(x))$ or $(x, y)$ as a real sample.

\textbf{Results}

Generally, segmentation performance of semi-cGAN or fewshot-GAN is slightly superior to 3D U-Net. The trained models are tested on both local dataset and cross dataset to investigate their accuracy. In particular,
\begin{itemize}
    \item\textbf{Local dataset:} the Dice score of semi-cGAN and fewshot-GAN are respectively $91.5117\%$ and $90.0431\%$, whereas the Dice score of 3D U-Net is $89.8137\%$. The average ASD of semi-cGAN and fewshot-GAN are 1.2726 and 1.5188, compared with 1.4747 of 3D U-Net.
    \item \textbf{Cross dataset:} similarly, the Dice score of semi-cGAN and fewshot-GAN are respectively $89.0644\%$ and $88.3881\%$, compared with $88.6382\%$ of 3D U-Net. The ASD of semi-cGAN and fewshot-GAN is 0.6869 and 0.6954, whereas the average ASD of 3D U-Net is 1.3109.
\end{itemize}

\section{Spinal Curvature Estimation Challenge 2019}

\textbf{Overview}
 
Scoliosis is a condition defined by an abnormal spinal curvature. In diagnosis and treatment planning of scoliosis, the Cobb angle is the most commonly used method employed for determining the degree of scoliosis. For instant, A Cobb angle which is greater than 45 degrees stands for a serious scoliosis. However, the manual measurement of Cobb angle is time consuming and unreliable; consequently, it is key to develop deep learning models that can automatically estimate Cobb angle. Due to the lack of effective methods that have been suggested for this purpose, the accurate automated spinal curvature estimation (AASCE) challenge aims to encourage and supply an objective evaluation platform for automatic spinal curvature estimation algorithms. 

\textbf{Task and dataset}

The task of this challenge is to automatically obtain 3 Cobb angles from each spinal anterior-posterior X-Ray image with a provided dataset which consists of a total of 707 spinal anterior-posterior x-ray images. The training set contains 609 images, and the testing set contains 98 ones. The landmarks were provided in the open training set, and the Cobb angles were computed adopting these landmarks. The ground truth of the testing set is only known by organizers. These annotations were determined manually by professional experts.

\textbf{Evalution and results}

The performance of participating methods are evaluated based on a symmetric mean absolute percentage (SMAPE). The SMAPE metric is defined as:

\begin{equation}
 SMAPE = \frac{1}{N}\sum_{i=1}^N\frac{SUM\left|X_i-Y_i\right|}{SUM\left|X_i+Y_i\right|}\ast100\%
\end{equation}
where $X_i$ is the estimated Cobb angles, $Y_i$ is the ground truth, $N$ is the number images.

After cleaning up some false and duplicate submissions, the 5 top-ranked results of the AASCE-2019 are displayed in Table \ref{table3.1}. Team X won this competition with $21.7135\%$ SMAPE.

\begin{table}[!htp]
\centering
\begin{tabular}{{l|l|l}}
\hline 
Ranking $(\#)$ & Team & SMAPE $(\%)$\\ 
\hline 
1 & X & 21.7135\\ 
\hline
2 & iFLYTEK & 22.1658\\
\hline
3 & Erasmus MC & 22.9631\\
\hline
4 & vipsl & 24.7987\\
\hline
5 & JLD & 25.4784 \\
\hline

 \end{tabular}
 \caption{Mean performance and ranking of the 5 top-ranked team on SMAPE.}
 \label{table3.1}
\end{table}

\subsection{Keypoint Detection for Spinal Curvature Estimation}

 Cobb angle has been the most useful material utilized for estimating the spinal curvature. In order to generate Cobb angles, spinal keypoints need to be detected initially. Accordingly, it is important to find a robust network that can perform precisely the detection task. As a result, K. Chen et al propose two methods~\cite{10.1007/978-3-030-39752-4_6} as below, which prove to be the state-of-the-art ones in detecting the spinal keypoints. 
 \begin{itemize}
     \item Method-1: A RetinaNet is firstly employed to predit the bounding box of each vertebra. A HR-Net is then applied to the output of RetinaNet to refine the keypoint detections.
     \item Method-2: A similar two-stage system is proposed. A Simple Baseline is initially used to extract 68 rough points along the spine curves employing. One then yields patches from that result and ensures that each patch includes three vertebrae at most with respect to ground truth. Following from that, a second Simple Baseline is trained to predict the exact keypoints of these patches, which are not fixed in numbers in general. Due to the freedom of patch selections, a delicate postprocess of clustering is implemented to handle dense keypoint predictions finally.
 \end{itemize}
 Research results show that by fusing the two methods above, they accomplish competitive outcomes With a symmetric mean absolute percentage error of $26.0535\%$ on the public leaderboard of AASCE-2019 challenge.

\textbf{ Methodology}

There are two methods that are proposed for the detection task:

\begin{center}
    \begin{figure}[htp]
    \begin{center}
     \includegraphics[scale=0.9]{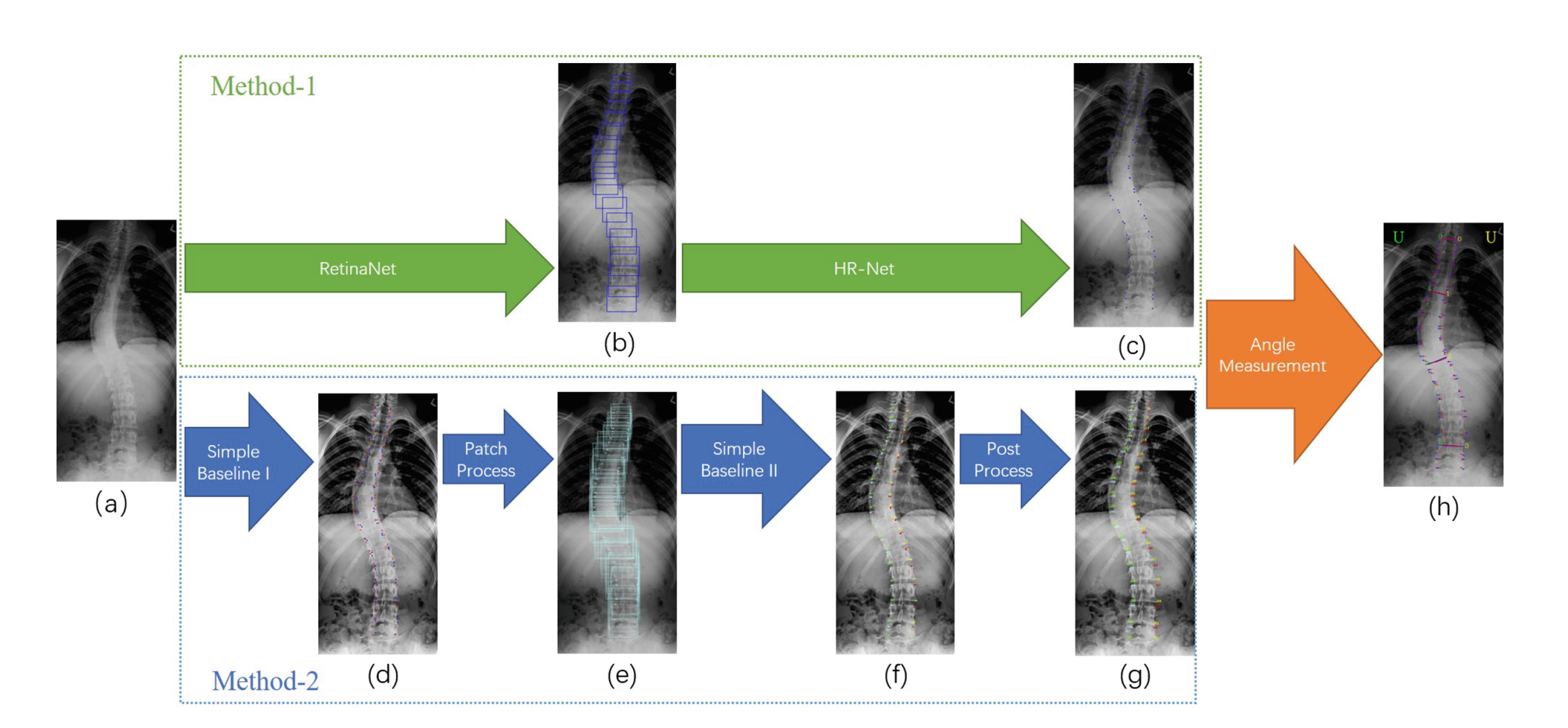}
    \end{center}
    \caption{Framework of the introduced two methods}
    \label{fig3.6}
    \end{figure}
\end{center}

\begin{itemize}
    \item \textbf{Method-1}: Method-1 utilizes a two-stage deep learning algorithm which is based on RetinaNet~\cite{8417976} and HR-Net~\cite{Fu_2022} (As see Fig. \ref{fig3.6}/Method-1). The first stage is to detect vertebrae and to yield corresponding bounding boxes employing RetinaNet. The RetinaNet is trained adopting the bounding boxes yielded from the 68 keypoints, requiring each box covers 4 keypoints of an individual vertebra. The second stage is to detect the 4 keypoints of the bounding box utilizing HR-Net.
    % detect vertebrae and then detect key points of each vertebrae!
    
    \item \textbf{Method-2}: Firstly Simple Baseline I~\cite{Xiao_2018_ECCV} is trained with spinal images to detect all 68 keypoints in a spinal sample. Simple Baseline I can grasp the global implicit sequentiality of keypoints through generating corresponding heatmaps simultaneously with fixed order. The predicted keypoints can smoothly trace the curvature of most spines. Although their landmarks are not precise enough to compute Cobb angles, in patch process, those keypoints are utilized as the outline sketches of spines to capture patches to force the model focus on local information in a certain range of vertebrae. A patch consists of n points, $\{n|n = 4, 6, 8, 10, 12\}$, for one to three vertebrae. Patches are yielded randomly in multiple times within a certain vertebrae range, and an image sample generates hundreds of patches. Next, Simple Baseline II is employed to detect keypoints from those patches. Finally, to sum up the final 68 keypoints from 4 vertex groups of points, a postprocess is proposed to cluster and remove outliers. And in Postprocess (i.e. from Fig.\ref{fig3.6}(f) to (g)), squeezed vertebrae assessment is optional.

\end{itemize}
Base on the clinical measurement methods, the spinal keypoints detected from those method are used to calculate the 3 Cobb angels that strongly support AIS (dolescent idiopathic scoliosis) assessment. 

The frameworks of the two methods are shown in Fig.\ref{fig3.6}:
\begin{itemize}
    \item The bottom workflow shows the Method-1, (a) is the input image, the blue bounding boxes in (b) and keypoints in (c) are prediction outputs from RetinaNet and HR-Net. 
    \item The bottom workflow presents the Method-2, consisting (d) rough points generation, (e) patch selection, (f) patch points generation and (g) clustering. (h) is the final image that presents the process of spine curvature estimation, with the red and blue keypoints deputizing ground truth and prediction respectively. The lines illustrates which vertebrae are chosen to compute the Cobb angles.
\end{itemize}

\textbf{Results}

To increase the accuracy of detecting spinal keypoints, the outcomes of the two methods above are fused to generate more competitive results. Specially, one result from Method-1 and two results from Method-2 are selected to fuse. Compared with simply averaging the angles - the common measurement method, the fusion
method reduce $1\%$ SMAPE (from the result of previous state-of-the-art model), reaching $22.16582\%$ finally.

\subsection{Vertebra Detector and Vertebra Corners Regression}

Cobb angle is widely used in the diagnosis and treatment of scoliosis. However, due to the time-consuming and unreliable results of existing Cobb angle measurement methods in clinical practice, B. Khanal et al. introduced an novel approach named 'Vertebra Detector and Vertebra Corners Regression'~\cite{10.1007/978-3-030-39752-4_9}, which automatically detects Cobb Angle from X-ray images. This automatic method has a creative structure that includes a vertebrae detector to firstly detect vertebrae as objects and a landmark detector which individually estimates the 4 landmark corners of each vertebra. Following from that, Cobb Angles are computed adopting the slope of each vertebra acquired from the predicted landmarks. To improve the prediction accuracy, they employ pre and post processing such as cropping, outlier rejection and smoothing of the predicted landmarks. They achieve SMAPE score of $25.69\%$ on the challenge test set, which indicates a promise.
% Nevertheless, manual measurement of Cobb angles is time-consuming, and the results are also heavily affected by the expert’s choice.

\textbf{Methodology}

\begin{center}
    \begin{figure}[htp]
    \begin{center}
     \includegraphics[scale=1.4]{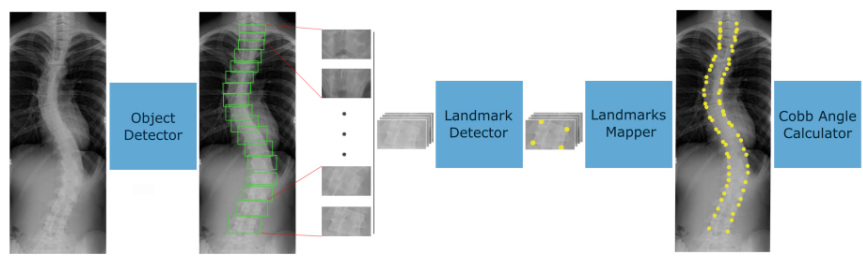}
    \end{center}
    \caption{ An overview of the pipeline}
    \label{fig3.7'}
    \end{figure}
\end{center}

The whole process is constructed by two networks (see Fig.\ref{fig3.7'}): one for vertebrae detection called object detector and one for landmarks regression called landmark detector. The architecture of object detection network is CNN-based Faster-RCNN~\cite{NIPS2015_14bfa6bb} while the landmark detection part employs  Densely Connected Convolutional Neural Network (DenseNet)~\cite{8099726}. Firstly, the input images are passed to an object detector to detect the vertebrae as bounding box objects. The detected vertebrae are then extracted as separate images and fed to a landmark detector to detect the four corners of the vertebra. Finally, the landmarks are mapped back to the original image from which Cobb angles are computed. 

As displayed in Fig.\ref{fig3.7'}, the object detector is utilized to detect the vertebrae as bounding box objects which are then passed to a landmark regression network as individual input images. The predicted normalized landmark coordinates from separate bounding boxes are associated and mapped back to the original images to compute Cobb angles.

\textbf{Results}

Detecting vertebrae as objects before predicting corner landmarks is found to be a promising approach. By applying pre and post processings that consist of cropping, outlier rejection and smoothing of the predicted landmarks, the best outcome results in a SMAPE score of $25.69\%$ on the challenge test set in AASCE challenge 2019, whereas the top score is $21.71 \%$. 

The point is that the proposed method does not properly consider the inter-dependency between landmark positions of various vertebrae. A learning algorithm to learn this inter-dependency would enhance the results.

\subsection{Spine Centerline Extraction with Ensembles of Cascaded Neural Networks}

As we know spinal curvature can be assessed using Cobb angles. However, existing Cobb angle measurement methods in clinical practice are time-consuming and unreliable. As a result, F. Dubost et al. propose an automated method for the direct estimation of Cobb angles from X-ray scans, which is named 'Automated Estimation of the Spinal Curvature via Spine Centerline Extraction with Ensembles of Cascaded Neural Networks'~\cite{10.1007/978-3-030-39752-4_10}. This method includes two main steps:
\begin{itemize}
    \item \textbf{step 1:} a cascade of two convolutional neural networks is utilized to segment the centerline of the spine.
    \item \textbf{step 2:} the centerline is smoothed and Cobb angles were automatically estimated employing the derivative of the centerline.
\end{itemize}

On the challenge’s test set, they acquired an average SMAPE of $22.96\%$

\textbf{Methodology}

There are two steps to directly predict 3 Cobb angles from the X-ray scans:
\begin{itemize}
    \item \textbf{Centerline Extraction with Neural Networks:} the centerline of the spine is extracted employing two cascaded convolutional neural networks whose architecture was a U-Net with fewer feature maps and batch normalization layers~\cite{pmlr-v37-ioffe15} before each pooling layer (see Fig. \ref{fig3.8}). The networks are simultaneously optimized end-to-end. The X-ray scan is used as an input for the first network which is optimized to segment the complete spine, while the second network utilized the output of the first network as input to segment the centerline only. Ground truth segmentations of the complete spine and the centerline were automatically calculated adopting the landmarks available in the training dataset.
    
     \begin{center}
    \begin{figure}[htp]
    \begin{center}
     \includegraphics[scale=0.9]{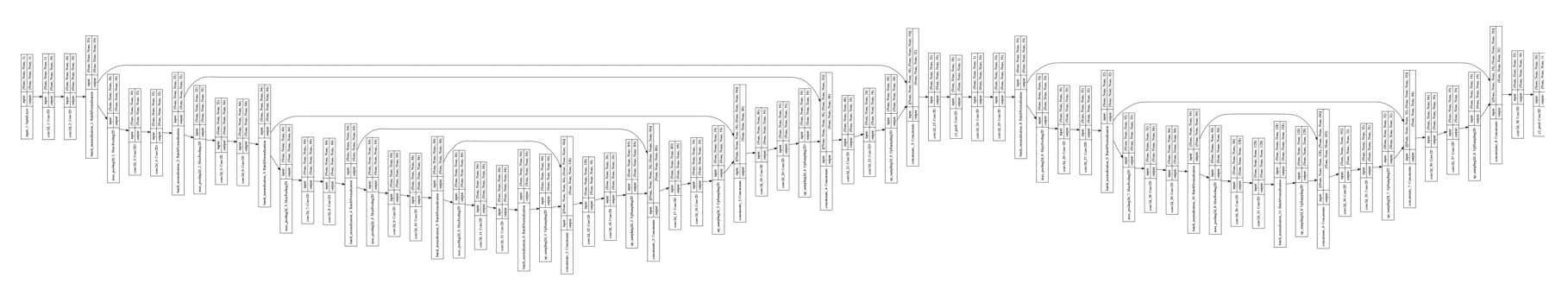}
    \end{center}
    \caption{Rough cascaded networks architecture}
    \label{fig3.8}
    \end{figure}
    \end{center}
    
    \item\textbf{Postprocessing:} the output of the second neural network is continued to go through the post-processing pipeline (displayed in Fig.\ref{fig3.8'}) including thresholding the centerline segmentation; removing small connected components; extracting the spine centerline curve; centerline smoothing using heat equation; computing the derivative of the centerline, which was subsequently used to compute Cobb angles.
    
    \begin{center}
    \begin{figure}[htp]
    \begin{center}
     \includegraphics[scale=0.8]{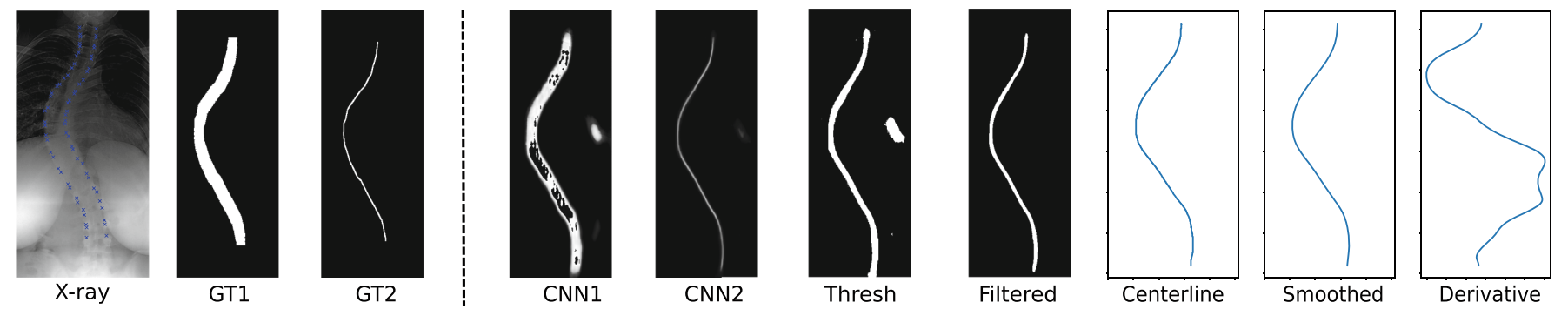}
    \end{center}
    \caption{Spine segmentation and curvature computation}
    \label{fig3.8'}
    \end{figure}
\end{center}
\end{itemize}

The summary of this approach is presented in Fig.\ref{fig3.8'}:

\begin{itemize}
    \item The first column from left to right consists of:
    \begin{itemize}
        \item  An input X-ray scan with corners of the vertebrae indicated manually in blue (Xray);
        \item Spine segmentation by connecting vertebral corners (GT1) and centerline (GT2), employed for training the cascaded networks;
    \end{itemize}
    \item CNN1 and CNN2 are the outputs of the cascaded networks respectively: the spine and spine centerline segmentations.
    \item The bottom row presents the post-processing pipeline which includes:
    \begin{itemize}
        \item Thresholding the centerline segmentation (Thresh);
        \item Temoving small connected components (Filtered);
        \item Extracting the spine centerline curve (Centerline);
        \item Centerline smoothing using heat equation (Smoothed);
        \item Calculating the derivative of the centerline (Derivative), which was subsequently adopted to calculate Cobb angles.
    \end{itemize}
\end{itemize}

\textbf{Results}

\begin{center}
    \begin{figure}[htp]
    \begin{center}
     \includegraphics[scale=1.3]{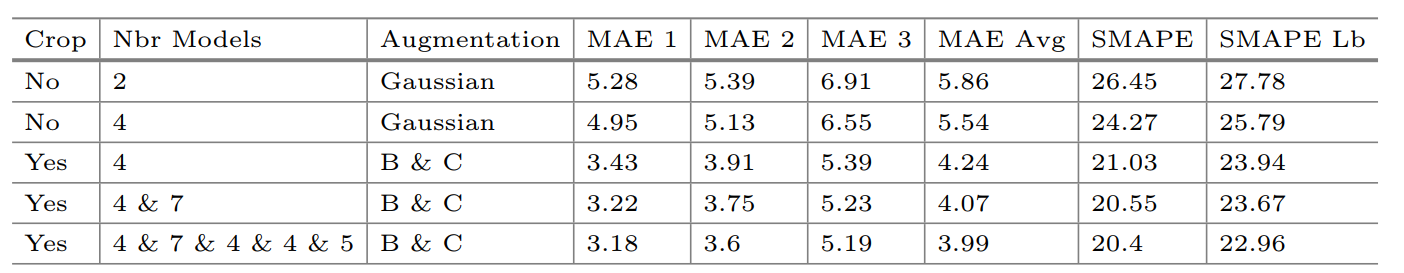}
    \end{center}
    \caption{Automated predictions of Cobb angles}
    \label{fig3.8''}
    \end{figure}
\end{center}

SMAPE and MAE (mean absolute error between the manual and automated assessment of Cobb angles) are computed to assess the prediction performance. All results are shown in Fig.\ref{fig3.8''}. On the challenge’s test set, this approach acquired a SMAPE of $22.96\%$

\subsection{Deep Heatmap Regression for Adolescent Idiopathic Scoliosis Assessment}

Cobb angle plays an core role in the diagnosis of scoliosis, which can quantify the degree of scoliosis effectively. Nevertheless, manual measurement of Cobb angles is time-consuming, and the results are also heavily affected by the expert's choice. In the challenge AASCE 2019, Z. Zhong et al. introduced an automatic detection framework for AIS assessment from X-ray CT images, which is named as their deep learning based coarse-to-fine heatmaps regression method~\cite{10.1007/978-3-030-39752-4_12}. Their result accomplishes an SMAPE of 24.7987 which is promising compared to other successful methods.

\textbf{Methodology}

\begin{center}
 \begin{figure}[htp]
    \begin{center}
     \includegraphics[scale=1.6]{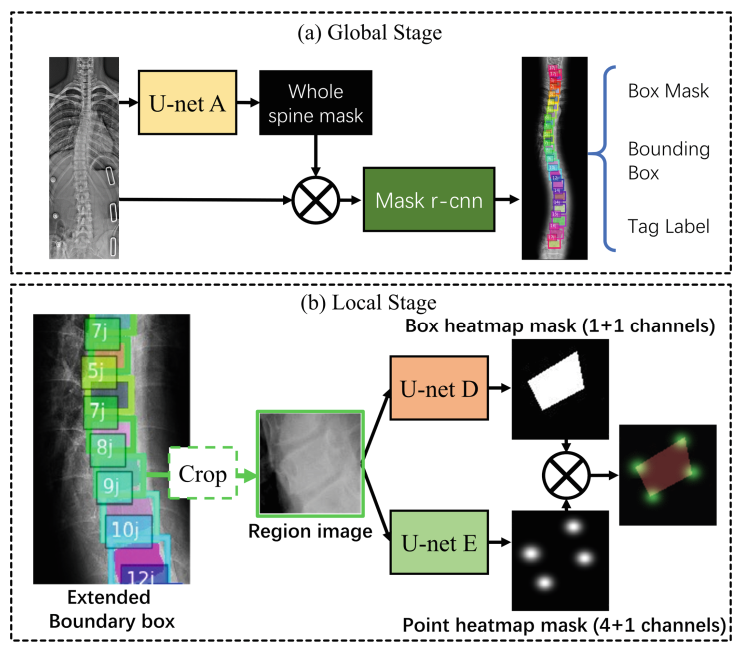}
    \end{center}
    \caption{The main parts of deep learning based coarse-to-fine heatmaps regression method framework}
    \label{fig3.9}
    \end{figure}
\end{center}

To calculate the three Cobb angles, the proposed method will first predict landmarks and then the Cobb angles are estimated adopting the detected vertebrate landmarks. As displayed in Fig.\ref{fig3.9}, the overall framework for landmark detection consists of 2 stages. 

\begin{itemize}
    \item \textbf{Global stage:} In the global stage, the U-net A~\cite{10.1007/978-3-319-24574-4_28} first takes the peeling cropped image as input to regress the whole spine mask. This mask is then associated with the original image to form inputs to Mask r-cnn~\cite{8237584}, whose outputs include box masks, bounding boxes and tag labels.
    \item \textbf{Local stage:} In the local stage, the bounding box is widened and we crop the region image that accommodated the box. The local region image is then passed separately to U-net D and U-net E, whose outputs are multiplied together. The landmarks of one vertebral column are the 4 highlights in the 4 channels heatmaps. The local stage procedure traverses the bounding boxes and yields the spine landmarks. 
\end{itemize} 
Based on this two-stage work, Z. Zhong et al. also recommended another landmark acquiring method. In particular, they make adjustments to the landmarks with polynomial curve fitting to modify the unusual landmarks and reduce the outliers; hence, raise the robustness.

\textbf{Results}

The data is supplied by the grand challenge AASCE 2019, and the results are the online evaluations on the unlabeled test dataset. These evaluation results are listed in Table \ref{table3.2}, and the proposed method demonstrates good performance and strong robustness. These results finally rank $4{th}$ place on  the 5 top-ranked team on SMAPE.

\begin{table}[!htp]
\centering
\begin{tabular}{{l|l}}
\hline 
Stage & SMAPE \\ 
\hline 
Local stage & 26.4455 \\ 
\hline 
Curve refine & 24.7987 \\ 
\hline
    \end{tabular}
    \caption{ The stage-wise results of the proposed framework.}
    \label{table3.2}
\end{table}

\subsection{Spinal Curvature Assessment Using Landmarks Estimation Network}

In the challenge AASCE 2019, R. Tao et al. suggested a novel two-stage method named 'Automated Spinal Curvature Assessment from X-Ray Images Using Landmarks Estimation Network via Rotation Proposals'~\cite{10.1007/978-3-030-39752-4_11}, which can directly predict Cobb angle from vertebrate landmarks. In this method, they employ rotation vertebrate region proposals to improve the precision of vertebrate localization in curved spinal region. Their network utilizes a backbone of ResNet50~\cite{He2016DeepRL} associated with Feature pyramid network (FPN) to extract multiscale region proposals. The rotation proposals are co-registered and inputed to stage-two fully convoluted network (FCN)~\cite{7298965} to detect vertebrate landmarks. The proposed method outperforms the traditional landmarks segmentation models for datasets with large variance. It finally achieves a SMAPE score of 25.4784 in the challenge AASCE 2019.

\textbf{Methodology}

 This work requires a two-stage automated spinal landmarks detection network based on rotational regional proposals of vertebrates. In the first stage, the suggested model detects location of vertebrates utilizing rotated rectangular regions. In the second stage, each vertebrate region firstly undergoes rotation co-registration employing rotation angle from the previous stage. Landmarks detection is then achieved from aligned proposals. And Cobb angle is finally computed adopting the detected landmarks.
 
\begin{center}
 \begin{figure}[htp]
    \begin{center}
     \includegraphics[scale=1.4]{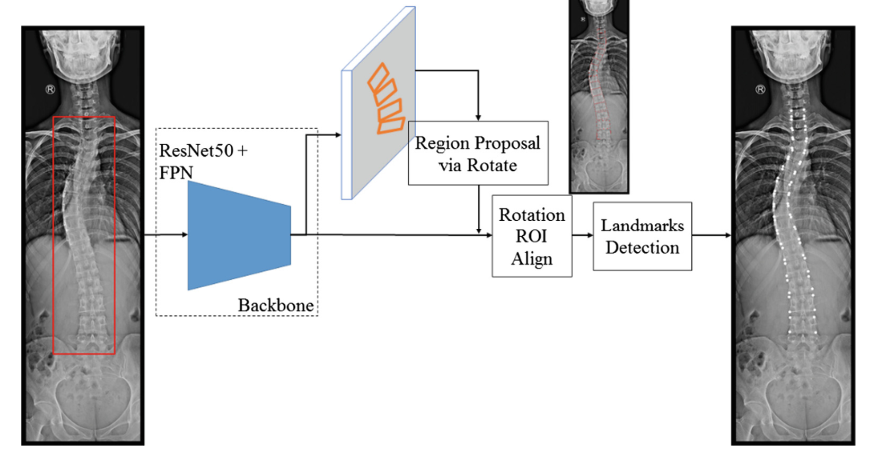}
    \end{center}
    \caption{Pipeline of proposed method}
    \label{fig3.10}
    \end{figure}
\end{center}

\begin{itemize}
    \item \textbf{Stage 1: Vertebrates Detection via Rotational Proposals} FPN is used as network’s backbone (see Fig.\ref{fig3.10}) extracting multiscale features and producing regional proposals on each scale. The input ground truth of each rectangular vertebrate bounding box contains 5 parameters $(x, y, w, h, \theta)$, where 
    \begin{itemize}
        \item  $(x, y)$ is bounding box center location;
        \item $(w, h)$ is bounding box dimension;
        \item $\theta$ the angle of tilted bounding box with respect to x-axis with rotation center fixed at $(x, y)$, ranging from $-\frac{\pi}{2}$ to $\frac{\pi}{2}$.
    \end{itemize}
Besides, rotational anchors are assigned at 5 scales, 3 ratios and 3 rotational angles. 
    \item \textbf{Stage 2: Rotational ROI Align and Landmarks Detection} As displayed in Fig.\ref{fig3.10}, Rotational region of interest (ROI) align is adopted on vertebrate proposals. Rotated vertebrate proposals are firstly modified employing rotation angles which are regressed from stage 1. These proposals then undergo ROI align into fixed size feature maps. Eventually, they are fed into fully convolutional network (FCN) to detect landmarks.
\end{itemize}

They use ResNet50 combined with FPN as backbone for multiscale feature extraction. Rotational region proposals are extracted from stage one and fed into FCN for landmarks detection.

\textbf{Results}

Despite significant difference among train, validation and test dataset, the proposed network accomplished promising results on the challenge test set. It is also compared with V-Net and Cascade Pyramid Network (CPN). Experiment results show that this method accomplishes better performance than the other two. Their SMAPE score are described in the table below:   

\begin{table}[!htp]
\centering
\begin{tabular}{{l|l|l|l}}
\hline 
Model & V-Net & CPN & proposed method \\ 
\hline 
SMAPE & 30.7135 & 40.7873 & 25.4784 \\
\hline
    \end{tabular}
    \caption{ Quantitative comparison of three different methods}
    \label{table3.3}
\end{table}

\chapter{Implementation of Seg4Reg Networks for automated Spinal Curvature Estimation}
\section{Overview}

Adolescent idiopathic scoliosis (AIS) is the most popular type of spinal deformity in children during early puberty. In the diagnosis and treatment of scoliosis, accurately estimating spinal curvature plays a significant role, and Cobb angle has been known as the most commonly employed method to determine the degree of scoliosis. However, existing Cobb angle measurement methods in clinical practice depend on doctors' manual intervention. Consequently, It is usually time-consuming and produces unreliable outcomes. Lately, we have seen that deep neural networks get considerable accomplishments in diverse image tasks such as classification, detection, regression etc. Therefore, applying deep models to the issue of spinal curvature estimation has become an attractive topic in automated AIS assessment. 
At the present, there are two approaches to estimate the Cobb angles: 
\begin{itemize}
    \item  Predicting landmarks and then angles~\cite{10.1007/978-3-319-66182-7_15,Wu2018AutomatedCA}: this strategy can generate high-precision angle outcomes but strongly depends on predicting landmarks. It implies a small error in coordinates may cause a big mistake in predicting angles. 
    \item  Regressing angle values~\cite{Chen2019AnAA}: The idea of this approach is to automatically estimate Cobb angle from vertebrate landmarks provided by professional doctors. These Angle regression methods are generally more stable but may lack the capacity to produce accurate predictions.
\end{itemize}
In practice, the experimental outcomes illustrate that the regression approach outperforms the landmark method. I will verify that by exploring and implementing a robust method named as Seg4Reg proposed by Y. Lin et al.,~\cite{10.1007/978-3-030-39752-4_7}, which ranks $1^{st}$ place on the 5 top-ranked teams of AASCE-2019. The main idea of their approach is to automatically estimate the 3 Cobb angles from X-ray scans with provided landmarks. I will present more details in next parts.

\section{Datasets}

\subsection{Training set}

\begin{center}
    \begin{figure}[htp]
    \begin{center}
     \includegraphics[scale=1.0]{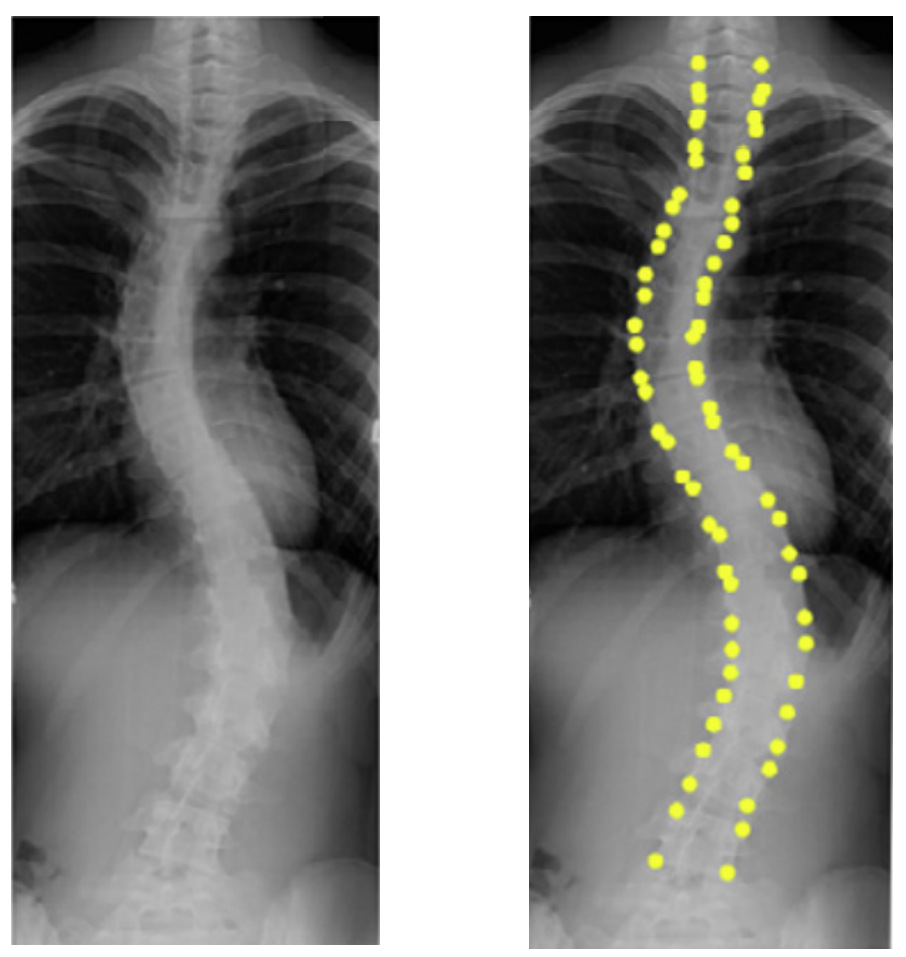}
    \end{center}
    \caption{ An image in the Training set and its landmarks}
    \label{fig4.1'}
    \end{figure}
\end{center}

The method was optimized with a publicly dataset of 609 spinal
anterior-posterior X-ray images available at SpineWeb\footnote{ http://spineweb.digitalimaginggroup.ca/spineweb/index.php?n=Main. Datasets.} as Dataset 16. It is collected from London Health Sciences Center in Canada employing EOS medical imaging system, and is utilized as the training set in the Accurate Automated Spinal Curvature Estimation (AASCE) challenge 2019. In the training process, it is splitted up into the train-set and test-set with 481 and 128 images respectively.

In each image, 17 vertebrae consisting of the thoracic and lumbar vertebrae were selected as part of the spinal curvature evaluation. Each vertebra is indicated by four landmarks at the four corners thereof (see Fig.\ref{fig4.1'}). There are totally 68 points per spinal image, and the Cobb angles were calculated using these landmarks. The landmarks were supplied by two professional doctors in London Health Sciences Center, and there are 3 Cobb angles in each image. 

\subsection{Testing set}

\begin{center}
    \begin{figure}[htp]
    \begin{center}
     \includegraphics[scale=1.5]{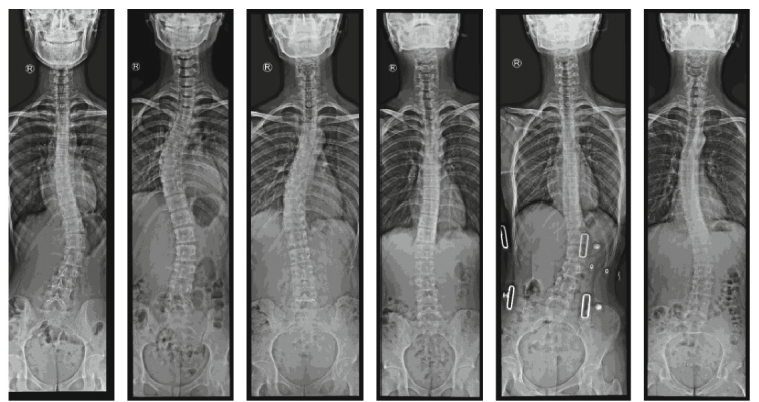}
    \end{center}
    \caption{Images in the Testing set}
    \label{fig4.2'}
    \end{figure}
\end{center}

The test dataset only released x-ray images with the ground truth only
known by organizers, which contains 98 X-ray scans. These images were different from the training images, which concentrated on the spine, in that they displayed areas of the neck and shoulders (see Fig.\ref{fig4.2'}). The images of the evaluation dataset will be manually cropped so that they illustrate the similar region as images of the training dataset. 

In addition, this dataset was substantially different from the training dataset in that the majority of its images had small Cobb angles (low curvature), while Cobb angles in the training set were more uniformly distributed. As additional preprocessing, the intensities of all images are normalized by dividing by the maximum intensity value of each image individually.

\section{Seg4Reg Network and Its Experimental Outcomes provided by Y. Lin et al. }

\subsection{Methodology}

\begin{center}
    \begin{figure}[htp]
    \begin{center}
     \includegraphics[scale=0.9]{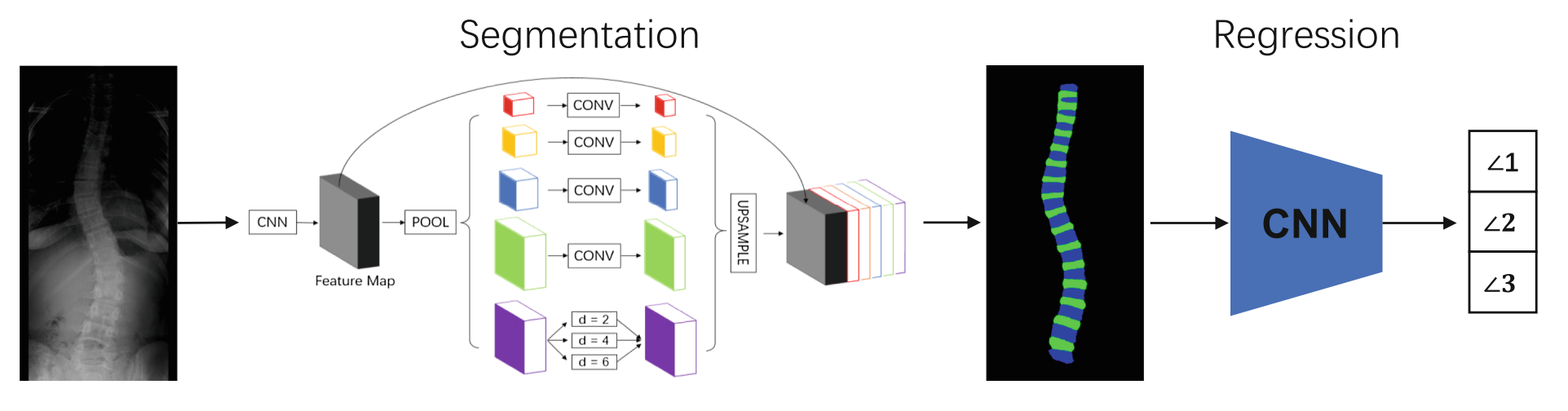}
    \end{center}
    \caption{ An overview of Seg4Reg's pipeline}
    \label{fig4.1}
    \end{figure}
\end{center}

Y. Lin et al. suggest a novel multi-task framework called Seg4Reg to generate precise spinal curvature estimation. This new approach proposes an automated method for the prediction of Cobb angles from X-ray scans. Its process contains two deep neural networks respectively concentrating on segmentation and regression ( see Fig.\ref{fig4.1}). In detail, the segmentation network first outputs the segmentation masks, and then based on these masks, the regression model directly predicts the Cobb angles. The architecture of segmentation model is similar to Pyramid scene parsing network (PSPNet)~\cite{Zhao_2017_CVPR} while the regression part makes use of traditional classification networks such as ResNet~\cite{7780459} or DenseNet. Besides, the domain shift issue which appeared between training and testing sets is mitigated by adding a domain adaptation module into the network structures. 

As displayed in Fig.\ref{fig4.1}, the X-ray is first processed by applying a segmentation network. Note that the groundtruth mask is formalized by utilizing the provided landmarks. The predicted mask is then passed to the regression model to execute angle value prediction.

\textbf{Preprocessing}

\begin{center}
    \begin{figure}[htp]
    \begin{center}
     \includegraphics[scale=0.9]{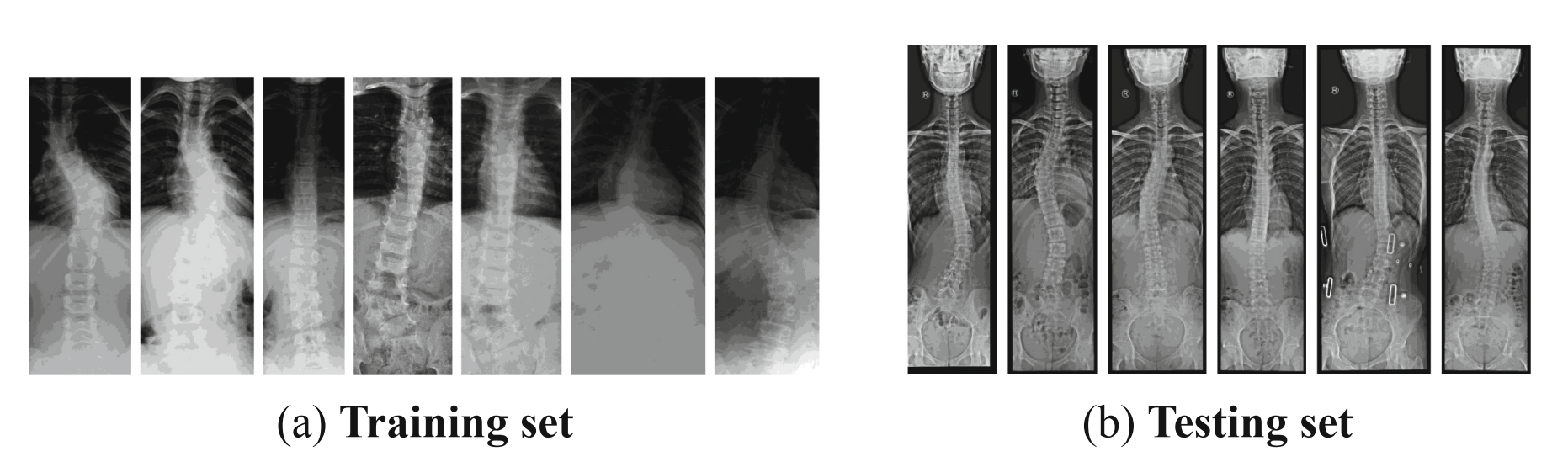}
    \end{center}
    \caption{A comparison of training set and testing set}
    \label{fig4.2}
    \end{figure}
\end{center}

To alleviate the explicit domain gap between training and testing sets shown in Fig.\ref{fig4.2}, histogram equalization is initially employed to both sets to make them similar visually. Moreover, due to the limited number of testing images, these X-rays are manually cropped to delete the skull and keep the spine in the appropriate scope. In addition, random rescaling $(0.85$ to $1.25)$ and random rotation $(- 45^{\circ}$ to $45^{\circ})$ are also applied during the training process for data argumentation. Gaussian noise is added to the input images as well to diminish the overfitting but it did not improve the prediction accuracy.

Besides, for the segmentation performance, the groundtruth masks are built on top of the offered landmarks' coordinates. In practise it is found that the segmentation model yields the best results by adding another class “gap between bones”. This operation may prevent overfitting in the training process that assists ultimate predictions more accurate.

\textbf{Network Architecture}

\begin{center}
    \begin{figure}[htp]
    \begin{center}
     \includegraphics[scale=1.3]{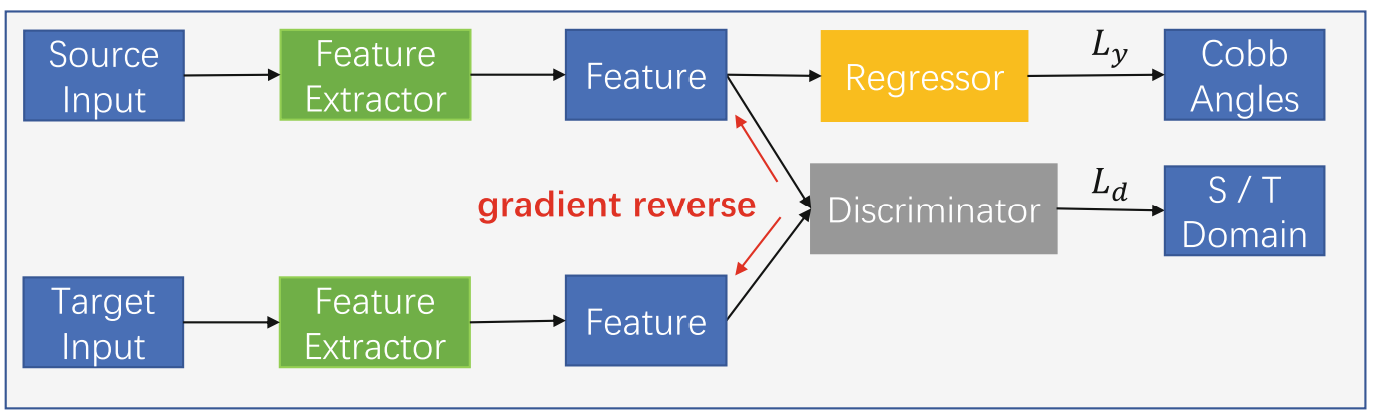}
    \end{center}
    \caption{The domain adaptation strategy}
    \label{fig4.3}
    \end{figure}
\end{center}

The suggestions in ~\cite{Zhao_2017_CVPR} are applied to create the segmentation model. As illustrated in Fig.\ref{fig4.1}, the input image is passed to a feature extractor CNN which generates feature maps initially. Next, the PSPNet ~\cite{Zhao_2017_CVPR} which uses various pooling kernels is applied to capture different receptive fields from the preceding output. Besides, the dilated convolution with various dilation rates such as 2, 4, 6 are appended to the pooling pyramid while summing their outputs after the convolution operations to remain the feature map size. Then commonly used feature extractors such as ResNet-50~\cite{7780459} and ResNet-101 are employed for backbone architecture (Fig.\ref{fig4.3}).

For the classification task, current classification models are directly utilized to do the regression part. Due to limited training samples, ImageNet based pretraining was employed as it assisted a lot. Additionally, the approach proposed in~\cite{Ganin2015UnsupervisedDA} was modified to solve the domain gap between training and testing sets by simply adding a discriminator branch and reversing its gradients during the back propagation. Therefore, the final loss function can be formalized as:
\begin{equation}
    Loss = L_y + \lambda L_d (1)
\end{equation}
where $\lambda$ is set to 1 in experiments.

\textbf{Network Training}

The default optimizer for both networks is Adam, and the initial learning rate is $3e^{-3}$. $0.9$ and $0.999$ are respectively allocated for $\beta_1$ and $\beta_2$. Y. Lin et al. also cosine annealing strategy and set weight decay to $1e^{-5}$. Each network was run for 50 epochs for the segmentation task and 90 epochs for the regression model. The segmentation input and the regression input were respectively resized to $1024\times512$ and $512\times256$. The ideal batch size is 32 on 4 NVIDIA P40 GPUs.

\begin{table}[!htp]
\centering
\begin{tabular}{{l|l|l|l|l}}
\hline 
Input type & Input size & Angle1 & Ange2 & Angle3 \\ 
\hline 
Img & (512, 256) & 6.0754 & 7.3386 & 6.7629 \\ 
\hline
Img + Mask & (512, 256) & 5.4489 & 6.4599 & 5.8470 \\
\hline
Mask & (512, 256) & $\textbf{4.7128}$ & $\textbf{5.7965}$ & $\textbf{5.6596}$ \\
\hline
Mask & (1024, 512) & 4.9360 & 7.2436 & 6.7321 \\
\hline
    \end{tabular}
    \caption{An ablative study on input types and input sizes.}
    \label{table4.1}
\end{table}

Table \ref{table4.1} shows a comparison on input types and input sizes. It is obvious that using segmentation mask as input achieves the best on the validation set while the best regression size is $(512, 256)$. The default segmentation backbone is PSPNet which appends the dilated convolution with different dilation rates to the pooling pyramid and regression model is DenseNet-169.

\subsection{Experimental Outcomes}

Y. Lin et al. presented experimental outcomes in both local validation and online testing sets. It is worth noting that cross validation was not utilized.

\textbf{Local Validation}

The L1 distance between model predictions and groundtruth labels is shown in Table \ref{table4.1}. It is easy to find that segmentation mask is the best input type and (512, 256) is the best input size. The comparison between the achievement of the improved version with PSPNet and DeepLab V3+~\cite{Chen_2018_ECCV} is displayed in Table \ref{table4.2}. It is clearly that the performance of previous PSPNet is improved by adding a dilation pyramid. 

\begin{table}[!htp]
\centering
\begin{tabular}{{l|l|l|l}}
\hline 
Metric & Y. Lin et al.'s & PSPNet & DeepLab V3+ \\ 
\hline 
mIOU & 0.8943 & 0.8715 & 0.817 \\ 
\hline
    \end{tabular}
    \caption{ The segmentation performance of PSPNet and DeepLab V3+}
    \label{table4.2}
\end{table}

\textbf{Online Testing}

The performance of the model is evaluated by competing in the AASCE challenge using the symmetric mean absolute percentage error (SMAPE). The online testing results are summarized in Table \ref{table4.3}:
\begin{itemize}
    \item The proposed dilation pyramid rises the online SMAPE by 0.48.
    \item EfficientNet-b5 is better than DenseNet-169 considering its higher ImageNet performance.
    \item The single model performance is then improved to 26.15 by adding domain adaptation module.
    \item  They improve the SMAPE score to 22.25 by ensembling the predictions of different models (they mainly ensembled ResNet series, DenseNet series and EfficientNet series).
    \item Since Angle2 and Angle3 are much smaller than Angle1 in the training set (Fig.\ref{fig4.6}), and Angle2 has many values which are close to zero, they reduced angles smaller than 4$^\circ$ to zeros which brought them to 21.71 SMAPE.
    \begin{center}
 \begin{figure}[htp]
 \begin{center}
\includegraphics[scale=0.9]{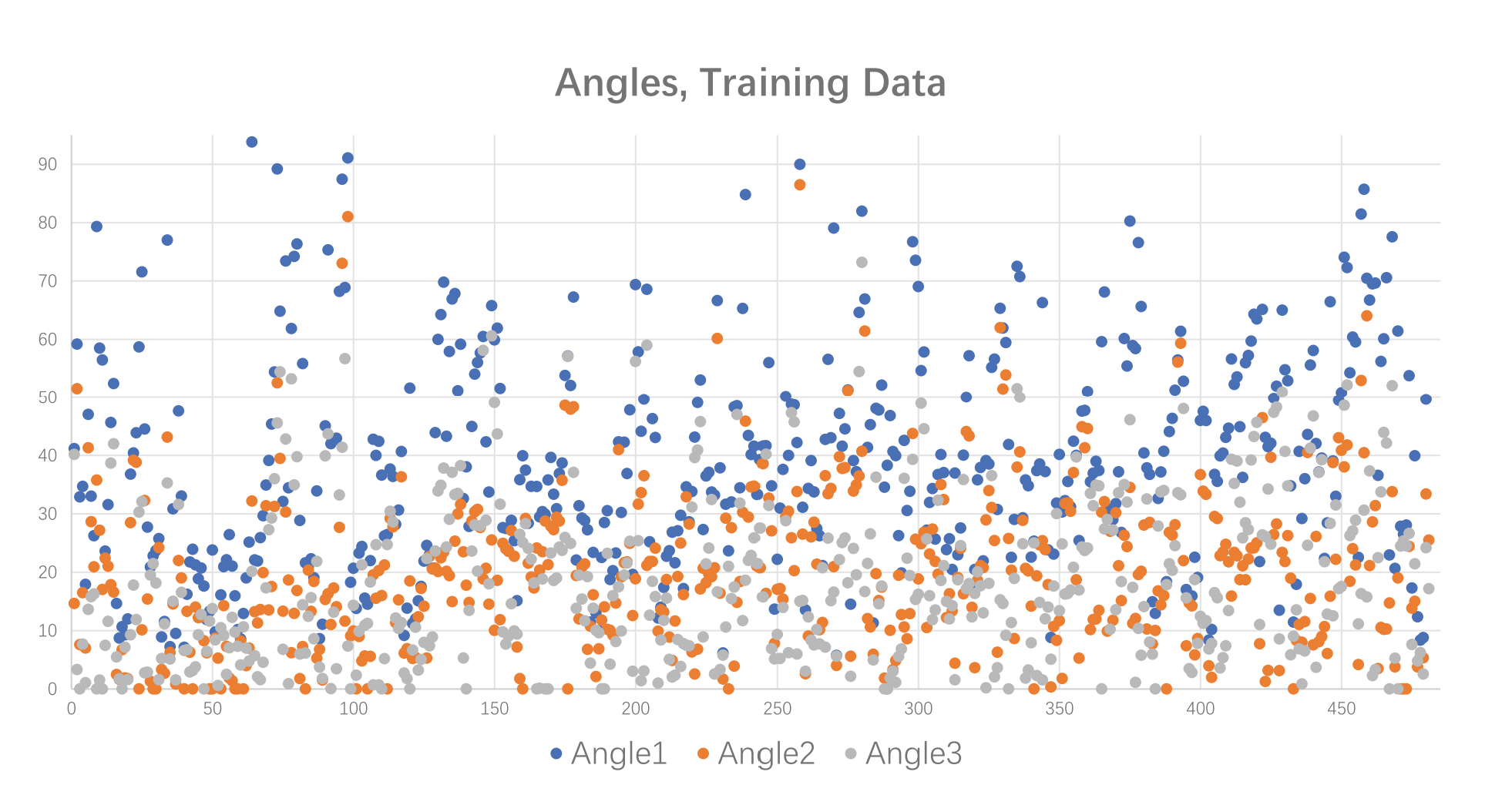}
\label{fig4.6}
\caption{Distribution of 3 angles in the training set}
\end{center} 
 \end{figure}
\end{center}
\end{itemize}

\begin{table}[!htp]
\centering
\begin{tabular}{{l|l|l|l|l|l}}
\hline 
Strategies &  &  &  &  & \\ 
\hline 
PSPNet + DenseNet-169 & \checkmark &  &  &   &\\ 
\hline
Y. Lin et al.'s + DenseNet-169 &  &  \checkmark &  &  &\\
\hline
Y. Lin et al.'s + EfficientNet-b5 &  &  &\checkmark&\checkmark & \\
\hline
Domain Adaptation &  &  &    &  \checkmark & \checkmark\\
\hline
Model Ensemble &  &  &  &   & \checkmark \\
\hline
SMAPE & 28.51 & 28.03 & 27.07 & 26.15 & 22.25\\
\hline
 \end{tabular}
 \caption{The Regression performances of various strategies}
 \label{table4.3}
\end{table}

\section{Implementation of Seg4Reg Network}

\subsection{Reused important concepts}
After analyzing the Seg4Reg Network of Y. Lin et al., I started to investigate their source code available on github~\cite{lin2019seg4reg}. My aim is to develop a better method that can directly output the Cobb angles from the input images with higher accuracy. The first idea is to improve the performance of their model by adjusting hyperparameters or change the backbone networks by some other robust networks etc. However, due to time constraints, my work ended at predicting Cobb Angle from the X-ray scans which consists of provided landmarks, using the Training set mentioned above.

I employed the preprocessing process, Network Architecture and Network Training supplied by Y. Lin et al. I skipped the segmentation part and used X-ray scans as direct inputs for the regression task. Traditional classification models such as DenseNets and ResNets are used as backbones of regression networks.

\subsection{Results}

The results of my implemetation is presented in Table \ref{table4.4}. Due to the limitation of GPU memory, I reduced the batch size of ResNet-152, DenseNet-161, DenseNet-169, DenseNet-201 to 16. DensNet-121 generated the best result with SMAPE of 23.9804 which is 2.2704 behind the best outcome of Y. Lin et al. in the AASCE challenge.

\begin{table}[!htp]
\centering
\begin{tabular}{{l|l|l|l|l|l}}
\hline 
Strategies & Batch size & SMAPE & Angle1 & Angle 2 & Angle 3\\ 
\hline 
ResNet 18 & 32 & 26.8208 &	21.0898	& 10.0208 &	14.4508\\ 
\hline
ResNet 50 & 32 & 26.1662 & 20.1404 & 10.0158 & 14.9147\\
\hline
ResNet 101 & 32 & 26.3562 &	21.4167 & 10.4560 &	14.3910\\
\hline
ResNet 152 & 16	& 28.8970 & 28.0550 & 11.3859& 17.3288\\
\hline
DenseNet 121 & \textbf{32} & \textbf{23.9804} &\textbf{17.7860} & \textbf{9.3920} & \textbf{11.9327} \\
\hline
DenseNet 161 & 16	& 27.2661 &	26.1598	& 9.8113 & 15.4096 \\
\hline
DenseNet 169 & 16 & 27.2661 & 26.1598	& 9.8113 & 15.4096 \\
\hline
DenseNet 201 & 16	& 27.6832 & 24.5159	& 12.7720 & 12.5473\\
\hline
 \end{tabular}
 \caption{The Regression performances of various networks.}
 \label{table4.4}
\end{table}

\subsection{Discussion}
In conclusion, I studied the Seg4Reg method and implemented the source code of it proposed by Y. Lin et al. My implementation adopts the main concepts of their pipeline but it is restricted to the regression task only. Though the result is good, it is still can not compete with the one they achieve at the online testing of the (AASCE) challenge 2019. It is because I did not apply earlier complex techniques such as segmentation, addition of domain adaption or ensemble modeling etc. due to time limitations. However, the experience which I received from studying and practising this state-of-the-art method gives me more hope for future improvement.

% body of thesis comes here

\chapter{Conclusion}

\label{ch:conclusions}

In this thesis, I presented the background theory of Deep Neural Networks, consisting of their main building blocks, commonly used models and their applications in Scoliosis Detection. Additionally, I reviewed the state-of-the-art deep neural network in 2019 Seg4Reg, which has accomplished the highest accuracy in the prediction of Cobb angle so far. Ultimately, implementation of this robust network helped me understand how Seg4Reg performed the best result in the Spinal Curvature Estimation Challenge 2019. A summary of most relevant thesis achievements, applications and potential future works are presented in the following sections.
\section{Summary of Thesis Achievements}

\subsection{Taking a general view on the relation of Applied Mathematics and Deep Learning}

From  an  applied  mathematics perspective, the familiar concepts such as calculus, approximation theory, optimization, and linear algebra form the basic ideas that underlie deep learning~\cite{higham2019deep}. Those are applied in both creating and training an deep neural network. In particular:
\begin{itemize}
    \item \textbf{Creating a network}. The core idea of a network starts with constructing a mapping that takes any 2D input (a spinal image) and returns the prediction outputs (e.g, estimation of 3 Cobb angles) by using repeated application of the nonlinear functions such as sigmoid, tanh, ReLU etc. 
    \item \textbf{Training a network}. The aim of the training process is to find the solution of an optimization problem. It includes (1) the stochastic gradient method or other optimization techniques that are designed to cope with very large scale sets of training data, (2) back propagation which described how partial derivatives are used for the optimization methods can be computed efficiently as well as (3) regularization techniques which is employed to improve the prediction accuracy.
\end{itemize}
Recently there has been a dramatic increase in the performance of image-based tasks due to the introduction of deep architectures. Yet, the mathematical explanations for this success continue to be elusive, and there is a need of mathematical justification for several properties of deep networks, such as global optimality, geometric stability, and invariance of the learned representations~\cite{vidal2017mathematics}. Thus, Applied Mathematics continues the task of verifying and improving the accuracy of DL-based models.
 
\subsection{The application Deep Learning in spinal imaging and Cobb angle prediction}

Spine related diseases or conditions are common and cause a huge burden of morbidity and cost to society. Spine imaging is an essential tool for assessing spinal pathologies, which is a process of utilizing deep neural networks to output expected prediction from spinal images. There have been a lot of research conducted to increase the performance of diverse spinal image task such as detection, regression, classification or segmentation.  Some of the popular application are presented as per below:
\begin{itemize}

    \item Detection of Vertebral Fractions, which is to predict vertebral compression fractures and thus, help prevent osteoporosis induced fractures.
    \item Vertebral Labelling in Radiographs, which is to localize and label vertebrae in spinal radiographs, and therefore, support the treatment planning of scoliosis and degenerative disorders.
    \item Segmentation of Multiple Lumbosacral Strutures, which is to segment nerve, bone, and disc from 3D images for spine related diseases in case of a few labeled data.
    \item Cobb angle estimation, which is to predict 3 Cobb angles from X-ray scans in order to support the diagnosis and treatment of scoliosis.
\end{itemize}

Since Adolescent idiopathic scoliosis (AIS) is one of the most common type of scoliosis, and Cobb angle is one of the most widely accepted standards to investigate the severity of scoliosis, the most attractive application of DL in spinal imaging is Cobb angle estimation. There have been increasing number of publications in recent years on this topic; however, there are two ways to estimate the Cobb angles at the present: (a) predicting landmarks and then angles~\cite{10.1007/978-3-319-66182-7_15,Wu2018AutomatedCA} and (b) regressing angle values~\cite{Chen2019AnAA}. The latest experimental results of Seg4Reg indicates that the regression strategy outperforms the landmark approach, and this state-of-the-art network achieves 21.71 SMAPE in the testing set of AASCE2019 challenge.

\subsection{The implementation of Seg4Reg network}

The implementation of this successful method is conducted using the main concepts provided by Y. Lin et al. Their pipiline contains two deep neural networks concentrating on segmentation and regression, respectively. Based on the results generated by the segmentation model, the regression network directly predicts the Cobb angles from segmentation masks. Nevertheless, my performance is limited to the regression part. It means that I estimate directly the 3 Cobb angles from the X-scans without segmenting them. Due to my limited time for this thesis, I also do not utilize other complex techniques to achieve the same final best results obtained by Y. Lin et al. However, by training various networks of DenseNets and ResNets, my work results in a promising outcome with a SMAPE score of 23.9804.

\section{Applications}

This thesis provides basic knowledge for those who are interested in both Applied Mathematics and Deep Learning as it explains how these two subjects contribute in spinal imaging. The main concepts of this thesis offer a general and specific background of how DL is adopted to address the challenge of spinal curvature estimation, and thus supply the understanding of Mathematics ideas behind it. Besides, people can build and train their own DL-based models base on the detailed presentation about key building blocks of deep neural networks, the commonly used deep neural networks and how to train them. Moreover, one can use this thesis as a template to implement the Seg4Reg for Spinal curvature estimation or other medical imaging tasks or they can develop more robust models from the popular ones to outperform thier achievements.

\section{Future Work}

Since Automated Spinal Curvature Estimation has been a challenging topic, there are still a lot of opportunities for improvement of the performance in the future. Among them, the two main future tasks that my thesis need to accomplish include:
\begin{itemize}
    \item Increase the precision of Seg4Reg network implementation by employing the segmentation task and improving its performance; adopting EfficientNet-b5 for regression part; applying domain adaption; utilizing model ensemble etc., which are proposed by Y. Lin et al.
    \item Enhancing the accuracy of Automated Spinal Curvature Estimation by adjusting the hyper parameters of Seg4Reg network; changing more practical backbone networks or even developing novel deep neural networks, which require more time and knowledge. 
\end{itemize}

\appendix
% appendices come here

\addcontentsline{toc}{chapter}{Bibliography}
\bibliographystyle{alpha}
\bibliography{bibliography}

\end{document}